\documentclass[10pt,letterpaper]{article}
\usepackage[top=0.85in,left=2.75in,footskip=0.75in]{geometry}

\usepackage{amsmath,amssymb}

\usepackage{changepage}

\usepackage[utf8x]{inputenc}

\usepackage{textcomp,marvosym}

\usepackage{cite}

\usepackage{nameref}
\usepackage{url}

\usepackage{microtype}
\DisableLigatures[f]{encoding = *, family = * }

\usepackage{array}

\newcolumntype{+}{!{\vrule width 2pt}}

\newlength\savedwidth

\raggedright
\usepackage[aboveskip=1pt,labelfont=bf,labelsep=period,justification=raggedright,singlelinecheck=off]{caption}

\makeatletter
\renewcommand{\@biblabel}[1]{\quad#1.}
\makeatother

\date{}

\usepackage{lastpage,fancyhdr,graphicx}
\usepackage{epstopdf}
\fancyheadoffset[L]{2.25in}
\fancyfootoffset[L]{2.25in}
\usepackage{verbatim}
\newcommand{\code}[1]{\texttt{#1}}
\newcommand{\BeatBox}{BeatBox}
\newcommand{\bbx}{BeatBox}
\newcommand{\bbg}{\code{.bbg}} 

\newenvironment{listing}{%
  \begingroup\footnotesize\verbatim%
}{%
  \endverbatim\endgroup%
}
\newcommand\pd{\partial}		
\providecommand{\Abs}[1]{\left\lvert#1\right\rvert}
\newcommand{\const}{\mathrm{const}}          
\renewcommand{\d}{{\mathrm d}}          
\newcommand{\Df}[2]{\frac{\d{#1}}{\d{#2}}} 
\newcommand{\df}[2]{\frac{\pd{#1}}{\pd{#2}}} 
\newcommand{\ddf}[2]{\frac{\pd^2{#1}}{\pd{#2}^2}} 
\newcommand{\e}{e}			
\newcommand{\eq}[1]{(\ref{eq:#1})}
\newcommand{\eqlabel}[1]{\label{eq:#1}}
\def\eqreftwo(#1,#2){(\ref{eq:#1},\ref{eq:#2})}
\newcommand{\eqtwo}[1]{\eqreftwo(#1)}
\newcommand{\etal}{\textit{et al.}}
\newcommand{\Fig}[1]{Fig~\ref{fig:#1}} 
\newcommand{\Figs}[1]{Figs~\ref{fig:#1}} 
\newcommand{\fig}[1]{Fig~\ref{fig:#1}} 
\newcommand{\figs}[1]{Figs~\ref{fig:#1}} 
\newcommand{\figref}[1]{\ref{fig:#1}}   
\newcommand{\figlabel}[1]{\label{fig:#1}}
\newcommand{\kron}{\delta}              

\newcommand{\mes}[1]{\mu\!\left(#1\right)}
\newcommand{\mx}[1]{\mathbf{#1}}        
\newcommand{\mV}{\mathrm{mV}}           
\providecommand{\Norm}[1]{\left\lVert#1\right\rVert}
\renewcommand{\O}[1]{\mathcal{O}\left(#1\right)}
\newcommand{\Real}{\mathbb{R}}          

\newcommand{\secn}[1]{Section~\secref{#1}}
\newcommand{\secref}[1]{``\nameref{sec:#1}''}
\newcommand{\seclabel}[1]{\label{sec:#1}}
\usepackage[tablename=Listing]{caption}
\newcommand{\Tab}[1]{Listing~\ref{tab:#1}} 
\newcommand{\tab}[1]{Listing~\ref{tab:#1}} 
\newcommand{\tablabel}[1]{\label{tab:#1}}
\newcommand{\Z}{\mathbb{Z}}             
\usepackage{pifont}
\newcommand{\+}[2]{\def#1{{#2}}}
\newcommand{\1}[2]{\def#1##1{{#2}}}

\+{\num}{^{\sharp}}           
\+{\ang}{\theta}              
\+{\BesselJ}{\mathrm{J}}      
\+{\Boundary}{\Gamma}         
\+{\C}{C}                     
\+{\CM}{C_m}                  
\+{\c}{c}                     
\+{\chf}{\psi}                
\+{\coeff}{\nu}               
\+{\D}{D}                     
\+{\Dten}{\hat{D}}            
\+{\Deten}{\hat{D}^e}         
\+{\Diten}{\hat{D}^i}         
\+{\Dpar}{D_{\parallel}}      
\+{\Dort}{D_{\perp}}          
\+{\Depar}{D^e_{\parallel}}   
\+{\Deort}{D^e_{\perp}}       
\+{\Dipar}{D^i_{\parallel}}   
\+{\Diort}{D^i_{\perp}}       
\+{\Deeff}{D^e_{*}}           
\+{\Dieff}{D^i_{*}}           
\+{\Deff}{D_{*}}              
\+{\Domain}{\mathcal{D}}      
\+{\di}{\Delta_i}             
\+{\dj}{\Delta_j}             
\+{\dk}{\Delta_k}             
\+{\dt}{k}                    
\+{\dx}{h}                    
\1{\E}{\times10^{#1}}         
\+{\err}{\varepsilon}         
\+{\f}{\mx{f}}                
\+{\fib}{f}                   
\+{\fibvec}{\vec{\fib}}       
\+{\fnb}{\beta}               
\+{\fng}{\gamma}              
\+{\fne}{\epsilon}            
\+{\g}{\mx{g}}                
\+{\Ie}{I_{\mathrm{ext}}}     
\+{\Ieff}{I_{\mathrm{eff}}}   
\+{\Iion}{I_{\mathrm{ion}}}   
\+{\Iu}{I_u}                  
\+{\Iv}{I_v}                  
\+{\i}{i}                     
\+{\j}{j}                     
\+{\jr}{\gamma}               
\+{\k}{k}                     
\+{\L}{\mathcal{L}}           
\+{\Len}{L}                   
\+{\Linf}{L^\infty}           
\+{\Ltwo}{L^2}                
\+{\n}{n}                     
\+{\normv}{\vec{n}}           
\+{\norm}{n}                  
\+{\r}{{\vec r}}              
\+{\sig}{\sigma}              
\+{\sigi}{\hat{\sig}_{\mathrm{i}}} 
\+{\sige}{\hat{\sig}_{\mathrm{e}}} 
\+{\sigeff}{\hat{\sig}_{\mathrm{eff}}} 
\+{\svr}{\chi}                
\+{\T}{T}                     
\+{\t}{t}                     
\+{\u}{u}                     
\+{\V}{V}                     
\+{\Va}{V^*}                  
\+{\v}{v}                     
\+{\weight}{W}                
\+{\p}{\mx{p}}                
\+{\phie}{\Phi_e}             
\+{\phii}{\Phi_i}             
\+{\phiia}{\Phi^*_i}          
\+{\pin}{_{\textrm{pin}}} 
\+{\ptv}{K}                   
\+{\q}{\mx{q}}                
\+{\s}{s}                     
\+{\x}{x}                     
\+{\xc}{x_*}                
\+{\y}{y}                     
\+{\yc}{y_*}                
\+{\z}{z}                     
\+{\zfa}{\alpha}              
\+{\A}{A}                     
\+{\B}{B}                     
\+{\alY}{\alpha}              
\+{\btY}{\beta}               
\1{\tC}{\tau_{#1}}            
\1{\yss}{\bar{y}_{#1}}        
\+{\vu}{\mx{u}}               
\+{\mA}{\mx{M}}               
\+{\mS}{\mx{S}}               
\+{\mD}{\mx{\Lambda}}         
\+{\mT}{\mx{T}}               
\+{\Ca}{c}

\begin{document}
\vspace*{0.2in}

\begin{flushleft}
{\Large
\textbf\newline{BeatBox --- HPC Simulation Environment for Biophysically and
  Anatomically Realistic Cardiac Electrophysiology} 
}
\newline
\\
Mario Antonioletti\textsuperscript{1},
Vadim N. Biktashev\textsuperscript{2,3},
Adrian Jackson\textsuperscript{1},
Sanjay R. Kharche\textsuperscript{2,4},
Tomas Stary\textsuperscript{2},
Irina V. Biktasheva\textsuperscript{5*,2},
\\
\bigskip
\textbf{1}  EPCC, The University of Edinburgh, Edinburgh, UK
\\
\textbf{2} CEMPS, University of Exeter, Exeter, UK
\\
\textbf{3} EPSRC Centre for Predictive Modelling in Healthcare,
  University of Exeter, Exeter, UK  
\\
\textbf{4} Institute of Cardiovascular Sciences, School of Medical Sciences,
University of Manchester, Manchester, UK
\\
\textbf{5} Dept of Computer Science, University of Liverpool, Liverpool, UK  
\\
\bigskip
%
%
* corresponding author: ivb@liv.ac.uk

\end{flushleft}
\section*{Abstract}
The \BeatBox\ simulation environment combines flexible script language
user interface with the robust computational tools,
in order
to setup cardiac
electrophysiology 
in-silico experiments without re-coding at low-level, so that cell
excitation, 
tissue/anatomy models, stimulation protocols may be included into a
\BeatBox\ script, 
and simulation run either sequentially or in parallel (MPI) without
re-compilation. 
\BeatBox\ is a free software written in C language to be run on a
Unix-based platform. 
It provides the whole spectrum of multi scale tissue modelling from
0-dimensional individual cell simulation, 
1-dimensional fibre, 2-dimensional sheet and 3-dimensional slab of
tissue, 
up to anatomically realistic whole heart simulations, with run time
measurements including cardiac re-entry tip/filament tracing, 
ECG, local/global samples of any variables, etc. BeatBox solvers, cell,
and 
tissue/anatomy models repositories are extended via robust and
flexible interfaces, 
thus providing an open framework for new developments in the field. 
In this paper we give an overview of the \BeatBox\ current state, 
together with 
a 
description of the main computational methods and MPI parallelisation approaches.


\section*{Introduction}


\subsection*{Background}
\seclabel{background}

Cardiovascular disease (CVD) is the main cause of death in Europe,
accounting for 47\% of all deaths \cite{bhf-cvd-2012}.  Cardiac
arrhythmias, where the electrical activity of the heart responsible
for its pumping action is disturbed, are among the most
serious CVDs. Despite over a century of study, the circumstances from which 
such fatal cardiac arrhythmias arise are still poorly
understood. Although several advancements have been made in linking
genetic mutations to arrhythmogenic CVD \cite{clancy:2008p2637,
  Noble:2002p777, Veldkamp:2000p2360}, these do not explain the resultant mechanisms by which
  arrhythmia and fibrillation emerge and sustain at the whole heart
  level, for the position of the heart in torso makes \emph{in vivo}
measurement awkward and invasive, prohibitively so for study in
humans. Thus, for some genetic cardiac diseases, the
  first presenting symptom is death with understandably limited opportunity to make even superficial
examinations \emph{in vivo}. The most modern experimental methods
  do not provide sufficient temporal and spatial resolution to trace
  down the multi-scale fine details of fibrillation development in samples of cardiac
  tissue, not to mention the heart in vivo. 

Combination of mathematical modelling
and the latest realistic
computer simulations of electrical activity in the heart have much
advanced our understanding of heart fibrillation and sudden cardiac
death \cite{%
      Clayton-etal-2011,%
      Zipes-Jalife-2014%
}, 
and the impact of \emph{in-silico} modelling, or indeed in-silico ``testing'', is expected to increase significantly as we approach the ultimate goal of the whole-heart modelling. 
With the vast amount of quantitative experimental data on cardiac myocytes action
potential and the underlying transmembrane ionic currents ready for
inclusion into the in-silico modeling, and the recent advance in high-resolution DT-MRI
provision of detail anatomy models, the biophysically and anatomically realistic
computer simulations allow
unimpeded access to the whole heart with greater spatial and temporal
resolutions than in a wet experiment, and allow to synthesise such elusive phenomena for
closer study, hence improving prospects of their treatment and
prevention. 

The biophysically and anatomically realistic simulation of cardiac action
potential propagation through the heart is computationally expensive
due to the huge number of equations per cell and the vast spatial and
temporal scales required. Complexity of realistic cardiac simulations
spans multi-physical scales to include greater detail at cellular
level, tissue heterogeneity, complex geometry and anisotropy of the
heart. Due to huge number of strongly nonlinear equations to be solved
on the vast temporal and spatial scales determined by the
high-resolution DT-MRI anatomy models, its timely running relies on
use of parallel processors - High Performance Computing (HPC). 

To address the intrinsically modular cardiac electrophysiology in silico
modelling, we developed modular software package BeatBox~\cite{%
  bbx-early,%
  VisionCSc-08,%
  McFarlane-2010%
}, with
a built-in simulation script interpreter, extendable repositories of cell
and tissue/anatomy models, capable to run both sequentially
and in parallel on distributed (MPI) memory architecture,~\fig{structure}. The Beatbox
cardiac simulation environment allows setup of complicated numerical experiments without re-coding at low-level, so that cell excitation, tissue and
anatomy models, stimulation protocols may be included into a script, 
and BeatBox simulation run either sequentially or in parallel without
re-compilation. Importantly, the BeatBox modular paradigm provides an
open framework for new developments in the field, for the open source
BeatBox solvers, and cell and tissue/anatomy repositories are extended
via robust and flexible interfaces. %
\begin{figure}[htb]
\centering\includegraphics{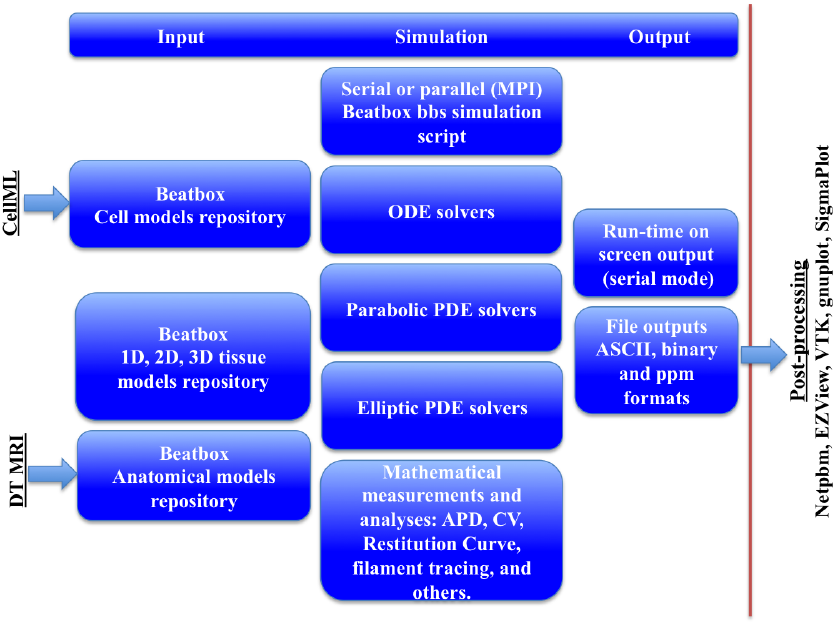}
\caption[]{\bf 
  \bbx\ formalism paradigm~\cite{VisionCSc-08}.
}
\figlabel{structure}
\end{figure}

 As HPC is a specialist field in its own right, it normally demands a high
qualification of the end users if they were to modify an existing
code for a different application problem. The main idea of \bbx\
is to make use of MPI routinely acessible to a widest community of users vast
majority of which are not professional software developers,
therefore \bbx's MPI
implementation aims to stay opaque to the user.

A number of successful computational cardiology
applications simulating electrophysiology and/or
biomechanics are available, e.g. %
CARP~\cite{CARP-2003}, %
CHASTE~\cite{CHASTE-2008}, %
Continuity~\cite{Continuity-2010}, %
CMISS~\cite{CMISS-2012}, %
Myokit~\cite{Myokit-2016}, 
CVRG Galaxy~\cite{GalaxyTools-Winslow-etal-2016}; %
further reviews and benchmark comparisons can found in %
\cite{Niederer-etal-2011-PTRSA}  %
and %
\cite{CardiacMechanicsBenchmark-PRSA-2015}. %
 Most of the available software is taylored for solution of limited
particular aspects of cardiac simulation.
A typical set-up allowed by electrophysiology simulation
software
is a certain number of electrical stimuli applied at certain times in
certain ways, and the user is allowed to vary the number and
parameters of these stimuli. Anything more complicated would require
re-coding at low level. As we show on particular examples in this
paper, the capabilities of \bbx\ workflow scripts are much wider. 
Another important feature is that \bbx\ finite
  difference implementation allows straightforward
  incorporation of Cartesian DT-MRI and/or micro-CT data into the cardiac
  electrophysiology simulations. Majority of the
  popular cardiac simulation software, including~\cite{%
    CARP-2003, %
    CHASTE-2008, %
    Continuity-2010, %
    CMISS-2012%
  }, use finite element discretization. 
  Conversion of anatomical data into finite element meshes is
  a separate step, requiring specialized tools, and can be a research task in
  itself~\cite{Plank-etal-Rabbit-2010}.

We believe that the
main achievement of \bbx\ modular and scripting approach is that it 
allows,
  on one hand, to
  maintain user flexibility for a large variety of simulation tasks, 
while on the other hand, to relieve the user from necessity
  of going into the code low level neither for changing the simulation
  protocol, nor for parallelisation detail. As the scripts contain
  no data specific to their use in parallel, the same scripts can
  run equivalently in both sequential and parallel modes (with the
  exception of run-time visualization).
In the subsequent sections, we outline the mathematical problems,
numerical methods and programming approaches characterising \bbx.

\subsection*{Cardiac Tissue Models}
\seclabel{tissue-models}

Computer simulation of cardiac muscle requires a mathematical model,
describing the relevant biophysical and electrophysiological
processes.  The \emph{bidomain} model
considers intracellular and extracellular spaces in the 
syncytium of cardiac myocytes. 
Those two domains are separated from each other by cellular
membranes, the conductivity through which is controlled by ionic channels.
This situation is described by a system of
partial (PDE) and ordinary (ODE) differential equations of the form: %
\begin{align} &
  \CM \df{\V}{\t} = - \Iion (\V,\g)  
  + \frac{1}{\svr} \nabla\cdot\sigi \nabla \phii, 
  \nonumber  \\ &
  \nabla\cdot(\sigi + \sige) \nabla\phii 
  = \nabla\cdot\sige \nabla\V - \Ie 
  \nonumber \\ &
  \df{\g}{\t} = \f(\g,\V,\r), 
          \eqlabel{bidomain}
\end{align}
where $\V$ is the transmembrane voltage, 
$\phii$ is the intracellular electrostatic potential
(so $\phie=\phii-\V$ is the extracellular potential), 
$\sigi$ and $\sige$ are the
anisotropic conductance tensors of the intra- and extracellular
domains respectively, $\CM$ is the specific capacitance of the membrane and
$\svr$ is the average surface to volume ratio of the cells.  The
transmembrane ionic currents $\Iion$ are controlled by gating
variables and ionic concentrations, represented by the vector
$\g$. The kinetic rates are expressed in terms of the vector-function
$\f$. The term $\Ie$ designates the external elecric current, say from experimental
or defibrillation electrodes.  In the system \eq{bidomain}, the first
equation is parabolic, the second is elliptic and the third
effectively is a system of ODEs at every point of the tissue
characterised by its location $\r$. If the intracellular conductances
are proportional, i.e. $\sige=\coeff\sigi$ for a scalar $\coeff$, 
then $\phii$, $\phie$ and $\V$ are proportional to each other, and
the system \eq{bidomain} simplifies to a \textit{monodomain} model: 
\begin{align} &
  \CM \df{\V}{\t} = - \Iion (\V,\g)
  + \frac{1}{\svr}\nabla\cdot\sigeff\nabla \V - \Ieff(\r,\t),  
  \nonumber \\ &
  \df{\g}{\t} = \f(\g,\V,\r), \eqlabel{monodomain}
\end{align}
where $\sigeff=\frac{\coeff}{1+\coeff}\sigi$,
$\Ieff=\frac{1}{\svr(1+\coeff)}\Ie$.
System~\eq{monodomain}
belongs to the class of
\textit{reaction-diffusion} systems, used for modelling of a large variety of 
natural and artificial nonlinear dissipative
systems \cite{Nicolis-deWit-2007}.

Computationally, the bidomain description is dramatically more
challenging than the monodomain, as the elliptic equation has to be
solved at every time step
(see e.g.~\cite{Trayanova-etal-2011}).
Practice shows that unless an external
electric field is involved, the bidomain models give results that
differ only slightly from corresponding appropriately chosen
monodomain models~\cite{%
  ColliFranzone-etal-2005,%
  Potse-etal-2006,%
  Nielsen-etal-2007,%
  Bishop-Plank-2011a,%
  Clayton-etal-2011%
}, which, together with the fact that experimental
data on the intra- and extracellular conductivity tensors are scarse,
means that in practice the monodomain simulations are used more widely.

The complexity of cardiac electrophysiology simulations 
further increases as it spans
multiple physical scales
to include greater detail at the cellular level, such
as cell signalling and metabolism, and greater integration with the surrounding
biological systems, such as electromechanical coupling and
vascular fluid 
dynamics~\cite{%
    Nash-Panfilov-2004,%
    Saucerman-McCulloch-2004,%
    Fink-etal-2011-PBMB,%
    Waters-etal-2011-PBMB,%
    Nordsletten-etal-2011-PBMB,%
    Zipes-Jalife-2014%
}.
In this context,
it is not surprising that the timely completion of simulations relies on 
modern high performance computing hardware.

Use of HPC facilities, although essential, is severely limited by
specialized software development skills required, so a separation of the
low-level coding from the processes of formulating and solving research problems
is highly desirable. The \bbx\ project seeks to overcome 
these difficulties by providing a computational environment that could
serve as a unifying paradigm for all \textit{in silico} cardiac electrophysiology research, and
for research in similar phenomena involving reaction-diffusion systems
outside the cardiology domain.

\section*{Design, Computational Algorithms, and Implementation}

\subsection*{ Logical structure and user interface}
\seclabel{interface}

The fundamental paradigm used by \bbx\ is to represent a simulation
as a ring of ``devices'', i.e. individual modules that
perform specific computational, input/output or control tasks. 
This ring is a metaphor of an iteration cycle;
  typically, one time step of calculations corresponds to one turn around
  the ring (see~\fig{ring}).
Module of each type can be used more than once in the ring, thus
providing more than one device instance. This ring of devices 
is constructed at  start-up,  based on the
instructions given in an
input script. 
The \bbx\ script parser places devices into the ring in the same order as
  they appear in the script.
The script describes the sequence of devices used in a
particular simulation and their parameters, using a domain-specfic
scripting language with a flexible syntax that includes things like a built-in
interpreter of arithmetic expression, recursive calls to other scripts,
etc. This allows \textbf{complicated numerical experiments
  to be set-up without low-level re-coding}. 

\begin{figure}[htb]
\centering\includegraphics{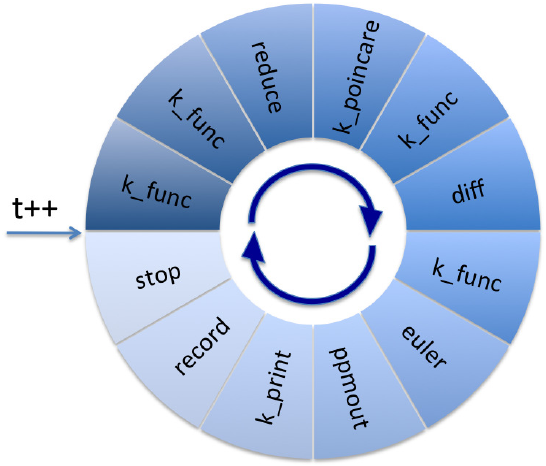}
\caption[]{%
{\bf \bbx\  ``Ring of devices''.} 
The ring of devices
  set up by \code{sample.bbs} script (see~\tab{sample} in the Appendix).
}
\figlabel{ring}
\end{figure}

With some simplification, \bbx\ computable data are of two
kinds: the main bulk of the data is in a four-dimensional
computational grid, of which three dimensions correspond to the spatial
dimensions, and the fourth dimension enumerates \emph{layers}
allocated for the components of the reaction-diffusion
system, \eq{bidomain} or \eq{monodomain}, as well as
for the output data produced by some of the
devices (e.g. those computing the
values of the diffusion term), or as a scratch memory area for 
devices that require this. When working with a complicated geometry, only
a subset of the regular cuboid spatial 2D or 3D 
grid corresponding to the tissue points is
involved. Every device typically works with only some layers of the
grid. Apart from the computational grid, there are global variables,
values of which may be used by some devices, and modified by 
others. 
From the MPI viewpoint, the main
 difference between
  the two types of data is that the
  computational grid is divided between threads, i.e. each thread has
  its own portion of the data, whereas the global
  variables are shared, i.e. each thread has the same set of global
  variables. The values of the global variables are identical in all
  threads; this is ensured either as they are produced as a result of
  identical computations in each thread, or, if a global variable is computed in
  one thread, its value is broadcast to other threads. 
An important feature is that any device instance in
the ring is associated with a global variable that serves as the flag
indicating whether this device instance should be executed at this particular
time step iteration in the computation. 

All MPI parallelization is done within
    individual devices, so conceptually, the device ring functions 
synchronously in all threads,
with actual synchronising \code{MPI\_Barrier} or \code{MPI\_Sendrecv}
    calls done only when required by the computation flow.
    Some devices operate on a sub-grid of the 4D data grid 
    (restricted ``space''), so that the set of active threads may
    change from device to device, and from one instance of a device to
    the other; if a device requires exchange of data between threads,
    then each device instance creates its own MPI communicator. 

The
MPI implementation of the script parsing does not present any
significant difficulties. The script file is read by all threads, and the threads allocate relevant subgrids of
the four-dimensional computational grid, corresponding to their
subdomains (see~\secn{partitioning} for detail), to every device
they create. Some care is required for diagnostic output,
e.g. normally \bbx\ echoes the parsed version of the script to the
log file. For normal output, this function is delegated to one
thread; for error messages, the thread that detected the error
will report it, which in case of massive parallelism may result in
a very large log file if the error occurs in many threads. %

Two simple examples of \bbx\ scripts are provided in
  the Appendix.  One of them is ``minimalist'', corresponding to a
  simulation protocol, that can be done by many other cardiac
  simulation softare. The second script is slightly more complicated, in order to
  illustrate the \bbx\ specific features on an example of a
  simulation of the feedback-controlled resonant drift
  algorithm as described in \cite{Biktashev-Holden-1994}, with 
  \fig{ring} illustrating the corresponding ring of the devices.
This script
involves the emulation
of an electrode registering electrical activity of the cardiac muscle
at a point, and of an external device which switches on a time-delay line
when a signal from the registering electrode satisfies a certain
condition, and issues a low-voltage
defibrillating shock upon the expiration of the time delay.
We stress here that implementation of this protocol in the
majority of other cardiac simulation software would require
modifications of the code at a low level.

We believe that the main features of \bbx\ are the
flexibility of its user interface, and the fact that any new computational features
can always be added as a new device. At the same time, it is clear that
its utility at present depends on specific computational
capabilities. In the next sections, we describe the
\bbx\ components 
most likely to be required in a typical cardiac dynamics
simulations, in their current state.
In the main text of the paper we focus on principal
  utility features of \bbx; for a more detailed description of the
  syntax and semantics of \bbx\ scripts, the reader is referred to the
  Appendix, whereas a fully comprehensive description is, of course,
  to be found in the user manual~\cite{beatbox-manual} which is distributed with the
  software and is not part of this communication.

\subsection*{Splitting the problem into parts}
\seclabel{divide-et-impera}

\subsubsection*{Computation of intermediate expressions}
\seclabel{intermediate}

``Divide and conquer'' is a popular and successful strategy for
evolution-type problems.  The idea is to split the right-hand sides of
complex evolution equations to simpler components, implement solvers
corresponding to each of these components, and then coordinate the
work of the solvers, so together they solve the whole problem. The
modular structure of a \bbx\ job makes this approach particularly easy
to implement.  One way of doing so is by computing different parts of
the right-hand side by different devices, and
then allow the time-stepping device to use the results of these computations.
We illustrate this by using a simple example, with reference
to the 
script \code{sample.bbs} presented in the \tab{sample}
in the Appendix and illustrated in~\fig{ring}.
Ignoring the effect of an external electric
field for now, the mathematical problem solved by this script is:
\begin{align} &
  \df{\u}{\t} = \frac{1}{\fne} \left( \u-\u^2/3 - \v\right)
                 + \nabla\Dten\nabla \u, 
  \nonumber \\ &
  \df{\v}{\t} = \fne\left( \u + \fnb - \fng \v\right).
                 \eqlabel{fhn}
\end{align}
The script implements a forward Euler timestep (see~\secn{explicit}
below) for system~\eq{fhn}, using two devices:
\begin{listing}
diff v0=[u] v1=[i] Dpar=D Dtrans=D/4 hx=hx;
euler v0=[u] v1=[v] ht=ht ode=fhncub 
  par={eps=eps bet=@[b] gam=gam Iu=@[i]};
\end{listing}
According to the definitions of string macros in the script, \tab{sample},
the macro \code{[u]} in this fragment expands to 0, that is the very first
layer of the grid allocated to the $\u$ field, 
\code{[v]} expands 1, standing for the second layer of the grid,
allocated to the $\v$ field, and \code{[i]} expands to 2, which is the third layer
of the grid, for the value of the diffusion term in~\eq{fhn}, 
$\nabla\cdot\Dten\nabla\u=\svr^{-1}\nabla\cdot\sigeff\nabla\u$.
So, the \code{diff} device computes an auxiliary variable
\begin{align*}
(\Iu)_\n = \nabla\cdot\Dten\nabla\u_\n
\end{align*}
where $\u_\n$ stands for the $\u$ field at the current time step
$\n$,
and stores it into layer \code{[v]=2} of the grid.

The \code{euler} device, with the parameter \code{ode=fhncub},
  performs a
forward Euler step for the cubic FitzHugh-Nagumo ODE system,
\begin{align} &
  \df{\u}{\t} = \frac{1}{\fne} \left( \u-\u^2/3 - \v\right) + \Iu, 
  \nonumber \\ &
  \df{\v}{\t} = \fne\left( \u + \fnb - \fng \v\right) + \Iv, 
  \eqlabel{FHN}
\end{align}
in which parameters $\fne$ and $\fng$ are given by the values of
the global variables \code{eps} and \code{gam} defined previously in the
script (in the included parameter file \code{<fhn.par>}),
the value of parameter $\fnb$ is taken from layer \code{[b]}
which expands to 3 (parameter $\fnb$ is spatially dependent in this
simulation, and layer 3 was pre-filled with values by the same
\code{k\_func} device that computed the initial conditions), the value
of parameter $\Iu$ is taken from layer $\code{[i]}$ which contains the
values of the anisotropic diffusion term $(\Iu)_\n$, computed for this time step
by the preceding \code{diff} device, and the value of parameter
$\Iv=0$ by default. Overall, with $\u_\n(\x,\y)$ and
$\v_\n(\x,\y)$ designating the fields $\u$ and $\v$ at the $\n$-th time step,
the pair of devices computes
\begin{align*} &
  \u_{\n+1}=\u_\n + \dt \left[ 
  \frac{1}{\fne} \left( \u_\n-\frac13\u_\n^2 - \v_\n\right) + \nabla\cdot\Dten\nabla\u_\n
  \right],
  \\ & 
   \v_{\n+1}=\v_\n + \dt \fne\left( \u_\n + \fnb(\x,\y) - \fng \v_\n\right),
\end{align*}
where $\dt$ is the time step, represented by the global variable \code{ht} 
in the \bbx\ script. 

\subsubsection*{Operator splitting}
\seclabel{splitting}

Operator splitting is another popular ``divide and conquer''
strategy~\cite{Strang-1968,Hundsdorfer-Verwer-2003,Clayton-etal-2011}. 
Slightly simplifying, one  can say that in
this approach, the right-hand sides still are split into
simpler parts, but now an evolution sub-step is done for each such part in
turn, as if this part was the whole right-hand side.  For example,
computation of kinetics and diffusion in the right-hand side
of~\eq{fhn} can be split into the kinetics part and diffusion part,
and then one device performs the diffusion substep, and another device
performs the kinetics substep. So, the \bbx\ script fragment
from~\secn{intermediate} can be modified as
\begin{listing}
diffstep v0=[u] v1=[i] Dpar=D Dtrans=D/4 hx=hx ht=ht;
euler v0=[u] v1=[v] ht=ht ode=fhncub 
  par={eps=eps bet=@[b] gam=gam};
\end{listing}
\noindent
  where the device \code{diffstep} computes the diffusion term, and does a
  forward Euler step with it, as if this was the only term in the
  equation. Combined with the fact that now in the \code{euler}
device the parameter \code{Iu} is not specified so it defaults to zero,
the given fragment of the script implements the following computation scheme: 
\begin{align*} &
  \u_{\n+1/2}=\u_\n + \dt \left[
  \nabla\cdot\Dten\nabla\u_\n
  \right],
  \\ & 
  \u_{\n+1}=\u_{\n+1/2} + \dt \left[ 
  \frac{1}{\fne} \left( \u_{\n+1/2}-\frac13u_{\n+1/2}^2 - \v_\n\right) 
  \right],
  \\ & 
   \v_{\n+1}=\v_\n + \dt \fne\left( \u_{\n+1/2} + \fnb(\x,\y) - \fng \v_\n\right).
\end{align*}
Once again, this is just a simple example illustrating how the \bbx\
paradigm naturally fits the idea of operator splitting. This of course
applies first of all to the simplest (Lie) splitting; more
sophisticated, higher-order operator splitting schemes could be
implemented at the \bbx\ script level, or on the device, i.e. a C-source code
level.

From the MPI viewpoint, both methods of splitting problems into
  parts do not present any issues, since they are implemented on the
  level of interaction of devices involved, and any parallelization
  work is done within the devices.

\subsection*{Kinetics solvers} 
\seclabel{kinetic}

\subsubsection*{Explicit solvers}
\seclabel{explicit}

Both the monodomain ``reaction-diffusion'' models of the form~\eq{monodomain} or
the more complicated bidomain models~\eq{bidomain} have equations with time
derivatives. Solving those equations in \bbx\ is done as if they were
ordinary differential equations, 
\begin{align} &
  \Df{\V}{\t} = - \frac{1}{\CM}\left(\Iion (\V,\g)- \Ieff(\r,\t)\right)
  \nonumber \\ &
  \Df{\g}{\t} = \f(\g,\V,\r), \eqlabel{ODE}
\end{align}
(depending on $\r$ as a parameter)
either with the value of the
diffusion term, computed by the corresponding diffusion device,
appearing in the voltage equation, or within the operator-splitting
paradigm, i.e. performing time sub-steps as if the model was restricted
to the ODEs representing the reaction terms, leaving the
space-dependent part of the model to be computed at alternative
sub-steps. 

The simplest and arguably most popular in practice solver for
ODEs is the first-order explicit (time-forward) scheme known as the \textbf{forward
Euler} scheme, which for a system of ODEs~\eq{ODE} means:
\begin{align} &
  \V_{\n+1} = \V_{\n} - \frac{\dt}{\CM}
  \left(\Iion (\V_\n,\g_\n)- \Ieff(\r,\t_\n)\right)
  \nonumber \\ &
  \g_{\n+1} = \g_{\n} + \dt \f(\g_\n,\V_\n,\r), \label{Euler}
\end{align}
where $\t_\n$ is the $\n$-th value in the time grid,
$\dt=\t_{\n+1}-\t_\n$ is the time step, and $\V_{\n}=\V(\r,\t_\n)$,
$\g_\n=\g(\r,\t_n)$.  This scheme is implemented in \bbx\ in the
\code{euler} device.

The Euler scheme's well known disadvantages are its low accuracy due to
only first-order approximation of the ODE, and, as any explicit
scheme, only conditional stability (see e.g.~\cite{RidgewayScott-2011}). 
The first disadvantage does not
usually play a crucial role in cardiac dynamics studies as the proven
accuracy of cardiac kinetics models themselves is not particularly
high. There is, however, \code{rk4} device in \bbx, implementing
\textbf{Runge-Kutta fourth-order scheme} for cases when accuracy is
essential, and other standard explicit solvers may be easily
implemented in a similar way. The stability consideration is more
significant as it severely limits the maximal allowable time step
$\dt$ in stiff models, hence making simulations costly.

\subsubsection*{Exponential solvers}
\seclabel{exponential}

The standard solution to the stability problem is, of course, using
implicit or semi-implicit schemes. The latter possibility is much more
popular as fully implicit approaches for nonlinear equations are
numerically challenging. Among the semi-implicit approaches available
in cardiac dynamics, the exponential scheme for ionic gates, known as
the \textbf{Rush-Larsen technique}~\cite{Rush-Larsen-1978}, is very popular. 
The idea is based on
the observation that in the models of ionic excitability, 
since the
seminal work by Hodgkin and Huxley~\cite{Hodgkin-Huxley-1952},
an important role is played by equations of the form:
\begin{align}
  \eqlabel{gate}
  \Df{\y}{\t} = \alY(\V) (1-\y) - \btY(\V) \y ,
\end{align}
where the dynamic variable $\y$, called the \emph{gating variable},
possibly in conjunction with other gating variables, determines
the permittivity of certain ionic currents. A convenient (even if not
biophysically precise) interpretation is that a channel is open if all
of the gates controlling that channel are open, and the variable $\y$ is
the probability for that gate to be open. Hence $\alY$ and
$\btY$ are transition probabilities per unit of time, of a closed
gate to open, or for an open gate to close, respectively. In
equation~\eq{gate} the transition probabilities depend on the current
value of the transmembrane voltage $\V$, as in the Hodgkin-Huxley
model; in more modern models gating variables of some channels may
depend on other dynamical variables, say the concentration of calcium
ions. The importance of the gating variables is that equations of the
type~\eq{gate} are often the stiffest in the whole cardiac excitation
model. The Rush-Larsen scheme in its simplest form can be obtained by
assuming that $\V$ does not change much during a time step,
$\t\in[\t_{\n},\t_{\n+1}]$, and replacing $\V(\t)$ with the constant
value $\V_n=\V(\t_n)$ turns \eq{gate} into a linear equation with
constant coefficients, the solution of which can be written in a
closed form, which gives:
\begin{align}
  \y_{\n+1} = \A(\V_\n) + \B(\V_\n) \y_{\n}
                                        \eqlabel{RL}
\end{align}
where
\begin{align} &
  \A(\V)=\exp\left(-(\alY(\V)+\btY(\V)) \dt \right) ,
\nonumber\\ &
   \B(\V) = 
  \frac{\alY(\V)}{\alY(\V)+\btY(\V)} 
  \left[ 1 - \A(\V) \right].
                                        \eqlabel{RLcoeff}
\end{align}
As far as equation~\eq{gate} is concerned, this scheme is
unconditionally stable, and gives an exact answer if $\V(\t)=\const$,
i.e. its first-order accuracy depends exclusively on the speed of
change of the transmembrane voltage $\V$. This scheme is
implemented in the
\bbx\ device \code{rushlarsen}. Naturally, this device requires a 
more detailed description of the excitable model than
\code{euler}: the gating variables $\y$ and their transition
rates $\A$, $\B$ need to be explicitly identified for
\code{rushlarsen} whereas \code{euler} only requires a definition of
the functions computing the right-hand sides of the dynamic equations, i.e.
$\Iion$ and $\g$.
The standard Rush-Larsen scheme can be modified to
  improve its accuracy; e.g~\cite{Sundnes-etal-2009} proposed a
  predictor-corrector version which provides a second order accuracy. 
  Implementation of this method would require a description of the
  cellular model in the same \code{ionic} format as used by
  \code{rushlarsen} device; however it is not yet implemented in the current
  version of \bbx.

Some modern cardiac excitation models use a Markov chain description of
the ionic channels. This description is based on the assumption that
an ionic channel can be in a finite number of discrete states, and
transitions between the states can happen with certain probability per
unit of time, which may depend on control variables, such as
transmembrane voltage $\V$ or calcium ion concentration $\Ca$. 
The time evolution of the
vector $\vu{}$ of the probabilities of the channel to be in each
particular state is described by the system of linear ODEs, known in
particular as \emph{Kolmogorov (forward) equations}, or the \emph{master
  equation}
\begin{align}
  \eqlabel{markov-generic}
	\Df{\vu}{\t} = \mA(\V,\Ca) \vu ,
\end{align}
where $\mA$ is the matrix of transition rates.
The extension of the
Rush-Larsen idea to this system was done
in~\cite{Stary-Biktashev-2015}. Assuming again that the control variables do not
change much within a time step and replacing them with a constant,
$\V(\t)=\V_\n$ and $\Ca(\t) = \Ca_\n$ for $\t\in[\t_{\n},\t_{\n+1}]$,
the system~\eq{markov-generic} is a system of homogeneous linear equations
with constant coefficients and its exact solutions can be explicitly
written. Assuming that $\mA$ is diagonalizable, the resulting
computational scheme can be written as:
\begin{align}
  \vu_{\n+1} = \mT(\V_\n, \Ca_\n) \vu_{\n},
\end{align}
where
\begin{align} 
  \mT(\V, \Ca) = \e^{\dt\mA(\V, \Ca)}=\mS(\V, \Ca)\,\e^{ \mD(\V, \Ca) \dt}\,\mS(\V, \Ca)^{-1}
\end{align}
and $\mS$ and $\mD$ are respectively the matrix of eigenvectors and
the diagonal matrix of eigenvalues of $\mA$. This \textbf{matrix
Rush-Larsen} scheme is also implemented in the device
\code{rushlarsen} mentioned earlier. 

Finding eigenvalues and eigenvectors for the diagonalisation and
computing exponentials are relatively time consuming operations. For
that reason the \code{rushlarsen} device does a \emph{tabulation}.
That is, for the case when the coefficients $\A$, $\B$ depend on $\V$
and matrices $\mT$ depend only on one control variable, e.g. $\V$ (are
\emph{``univariate''}), their values are precomputed for a
sufficiently fine grid of the control variable at the start 
time~\cite{tabulation}.

If matrix $\mA$ depends on multiple control variables, e.g. both $\V$
and $\Ca$ (are \emph{``multivariate''}), it can sometimes be presented
as a sum of univariate matrices. Then \code{rushlarsen} uses Lie
operator splitting and integrates each of the subsystems associated
with each of the univariate matrices using the tabulated ``matrix
Rush-Larsen'' separately. For some kinetics models,
$\mA$ can be presented as a
sum of one or more univariate matrices and a remainder, which is
multivariate but uniformly small. In that case the subsystem
associated with the small remainder is done using the forward Euler
method. Finally, if any such decomposition is not possible, ``matrix
Rush-Larsen'' step still can be done, just without tabulation, but by
doing the diagonalization ``on the fly'', i.e. at the run time rather than
start time. Although such computation is relatively costly, the
benefit of larger time step may still outweigh the expense. The
possibility of tabulating multivariate function theoretically exists
but is not considered in \bbx\ due to resource implications.

The diagonalization is done using appropriate routines
from GSL~\cite{GSL}; the relevant subset of GSL is included in the \bbx\
distribution for portability and the users convenience.

Other methods of extending Rush-Larsen idea to Markov
  chains have been proposed; e.g. the ``uniformization method''~
  \cite{Gomes-etal-2015} based on computing partial
  Taylor series of the matrix exponential for the suitably
  preconditioned matrix~$\mA$. This method does not require finding
  eigenvectors, but the amount of computations depends on the required
  accuracy. 

From the MPI viewpoint, all kinetic solvers work on individual
  grid points, so parallelization does not present any issues.

\subsubsection*{Cell models}

  The current version of \bbx\ is provided with a library of cell
  models. The definitions of the models come in two different formats,
  called~\code{rhs} and \code{ionic} in \bbx\ language,
  for the two different classes of kinetic solvers described
  above. 

  The \code{rhs} format is used by the generic solvers such as Euler and
  Runge-Kutta. In this format, the corresponding \code{C} module defines
  a function that computes the vector of the time-derivatives of the
  dynamic variables, for a given vector of the current values of those
  variables.

  A practical amendment to this idealized scheme came from
  the necessity to incorporate models which are not easily presentable
  as systems of ordinary differential equations. This includes the
  models where the description of intracellular calcium buffers is in
  terms of finite rather than differential equations, and 
  also the models with the description of calcium-induced calcium release in terms of a delay
  with respect to the voltage upstroke
  inflexion point, as in the Luo-Rudy family of models. The descriptions
  of such models used in cardiac modelling practice is often in the
  form of a function that performs the time-stepping, rather than
  defines the right-hand sides of the ODE system. Hence, the \code{rhs}
  format allows the model defining function to also directly modify
  the state vector, and correspondingly have the time step as one of
  the parameters. Currently, \bbx\ has
  \code{rhs} definitions of the ``conceptual'' excitable models, such
  as
  FitzHugh-Nagumo~\cite{FitzHugh-1961,Nagumo-etal-1962,Winfree-1991},
  Barkley~\cite{Barkley-1991},
  and complex Ginzburg-Landau equation~\cite{Aranson-Kramer-2002}, and
  specifically cardiac models, such as 
  Fenton-Karma~\cite{Fenton-Karma-1998}, Luo-Rudy
  ``LRd''~\cite{Luo-Rudy-1994}, Courtemanche \etal\
  1998~\cite{Courtemanche-etal-1998}, ten Tuschher \etal\
  2004~\cite{tenTuschher-etal-2004} and ten Tuschher-Panfilov
  2006~\cite{tenTuschher-Panfilov-2006}.

  The \code{ionic} format is suitable for solvers explicitly exploiting the
  specific structure of cardiac and neural excitation models, currently implemented in
  \code{rushlarsen}. This solver
  handles both the classical Rush-Larsen scheme, and its matrix
  modification described above. The difference from the \code{rhs}
  format is that the vector of dynamic variables is split into a part
  that corresponds to gating variables, Markov chain variables, and
  ``other'', i.e. non-gating variables. Correspondingly, a module
  definining an \code{ionic} model is expected to export separate
  functions computing the transition rates for the gating and Markov
  variables, and ODE right-hand sides for non-gating variables. The current
    version of  \bbx\ has \code{ionic} definitions of
  the following models: Beeler-Reuter~\cite{Beeler-Reuter-1977},
  Courtemanche \etal\ 1998~\cite{Courtemanche-etal-1998}, ten
  Tuschher-Panfilov 2006~\cite{tenTuschher-Panfilov-2006} and
  Hodgkin-Huxley~\cite{Hodgkin-Huxley-1952}. 

  Both \code{rhs} and \code{ionic} libraries of cell models can be extended by adding
  new models in the appropriate format.
  Instructions for that, with examples, are
  provided in the \bbx\ documentation~\cite{beatbox-manual}. This includes an example of
  a semi-automatic conversion of a model from the \code{CELLML}
  format into a \bbx\ \code{rhs} module. Making such conversion
  completely automatic is entirely feasible and is one of the
  planned directions of \bbx\ development. Conversion to
  \code{ionic} format is more complicated as it requires
    syntactic distinction of gate and Markov chain variables and
    their transition rates from other variables, currently not
    available in the \code{CELLML} standard. %

\subsection*{Monodomain diffusion and boundary conditions}
\seclabel{monodomain}

We focus here on the device \code{diff} which is the main device
implementing the diffusion term in the monodomain diffusion, which
mathematically can be described as:
\begin{equation}
 \L\u = 
 \sum\limits_{\j,\k=1}^{3} 
 \df{}{\x_\j} 
 \left(
   \D_{\j\k}(\r)\,\df{\u}{\x_\j} 
 \right)                                \eqlabel{Lu}
\end{equation}
with the naturally associated non-flux boundary conditions, 
\begin{equation}
 \sum\limits_{\j,\k=1}^{3} 
 \n_\k \D_{\j\k}(\r)\,\df{\u}{\x_\j} 
  =0                                    \eqlabel{nonflux}
\end{equation}
where $\normv=\left(\norm_\k\right)$ is the normal to the boundary
$\Boundary$ of the domain $\Domain$, i.e. excitable tissue. 
Currently this is
implemented for the transversely isotropic case, i.e. when
the diffusion tensor
$\Dten=\left(\CM\svr\right)^{-1}\sigeff=\left(\D_{\j\k}\right)$ has
only two different eigenvalues: the bigger, simple eigenvalue $\Dpar$
corresponding to the direction along the tissue fibers, and the
smaller, double eigenvalue $\Dort$, corresponding to the directions
across the fibres, as this is the most popular case in modelling
anisotropic cardiac tissue (the modification for the general
orthotropic case is straightforward though). In this case,
\begin{equation}
  \D_{\j\k} = \Dort\kron_{\j\k} +
  \left(\Dpar-\Dort\right)\fib_\j\fib_\k,
                                        \eqlabel{sigma}
\end{equation}
where $\fibvec=\left(\fib_\k\right)$ is the unit vector of the fiber
direction. The simple finite-difference approximation of \eqtwo{Lu,nonflux}
in \code{diff} device is along the lines described, e.g. in
\cite{Clayton-Panfilov-2007-PBMB}. In detail, we have
\begin{align}
  (\L\u)_\p = \sum\limits_{\q\in\{0,\pm1\}^3}
  \weight^{\p}_{\q} \u_{\p+\q},
\end{align}
where $\p\in\Z^3$ is the 3D index of a grid node with position vector
$\r_\p$, 
$\u_\p=\u(\r_\p)$ is the value of the field $\u$ at the grid node $\p$,
$(\L\u)_\p$ is the value of the diffusion operator approximation at
that point, $\q\in\left\{0,\pm1\right\}^3$ is the grid node index increment, and the weights
$\weight^{\p}_{\q}$ are defined by the following expressions:
\begin{align}
  \weight^{\p}_{\q} = \overline\weight^{\p}_{\q} + \widetilde\weight^{\p}_{\q},
\end{align}
\begin{align}
  \overline\weight^{\p}_{\q}=\frac{\chf_{\p+\q}}{\dx^2}\begin{cases}
    \D^{\p}_{\j\j} , & \q=\pm\q_j, \\
    \frac12\D^{\p}_{\j\k} ,  & \q=\pm(\q_j+\q_k), \; \j\ne\k, \\
    -\frac12\D^{\p}_{\j\k} , & \q=\pm(\q_j-\q_k), \; \j\ne\k, \\
    0,                       & \q=\pm\q_1\pm\q_2\pm\q_3, \\
  \end{cases},
\end{align}
\begin{equation}
  \widetilde\weight^{\p}_{\q}
  = \frac{\chf_{\p+\q}\chf_{\p-\q}}{4\dx^2} \begin{cases}
    \pm \sum\limits_{\j=1}^3 \left(
     \D_{\j\k}^{\p+\q_\j} - \D_{\j\k}^{\p-\q_\j}
   \right), & \q=\pm\q_\k, \\
   0, & \textrm{otherwise,}
 \end{cases}
\end{equation}
\begin{align}
   \weight^{\p}_{(0,0,0)}= - \sum\limits_{\q\ne(0,0,0)}\weight^{\p}_{\q}, 
\end{align}
where $\j,\k\in(1,2,3)$, 
$\chf_\p$ is the grid indicator function of the domain
$\Domain$, that is, $\chf_\p=1$ if $\r_\p\in\Domain$ and $\chf_{\p}=0$
otherwise, $\D^{\p}_{\j\k}=\D_{\j\k}\left(\r_\p\right)$, 
$\q_1=(1,0,0)$, $\q_2=(0,1,0)$, $\q_3=(0,0,1)$,
and $\dx$ is the space discretization step.

The above discretization is probably the simplest possible approach;
there are alternatives available, see for example~\cite{Fenton-etal-2005-C},
however these require extra 
information about $\Domain$
beyond the grid function $\chf_\p$. We assess the approximation
properties of the simple scheme described above by solving the
following initial-boundary value problem for the diffusion equation:
\begin{align}
  \df{\u}{\t} &= \ddf{\u}{\x}+\ddf{\u}{\y}, \quad (\x,\y)\in \Domain, \; \t\in[0,\T]; 
                                        \eqlabel{HTestPDE} \\
  \u &=\BesselJ_0\left(\jr \sqrt{(\x-\x_0)^2+(\y-\y_0)^2}\right), \quad (\x,\y)\in \Domain, \; \t=0;
                                        \eqlabel{HTestIC} \\
  \df{\u}{\normv} &= 0, \quad (\x,\y) \in \Boundary, \; \t\in\Real_+; 
                                        \eqlabel{HTestBC} \\
  \Domain &= \left\{ (\x,\y) | (\x-\x_0)^2+(\y-\y_0)^2 \le 1 \right\},
                                        \eqlabel{HTestDomain} \\
  \Boundary &= \pd\Domain = \left\{ (\x,\y) | (\x-\x_0)^2+(\y-\y_0)^2 = 1 \right\}.
                                        \eqlabel{HTestBoundary}
\end{align}
Here $\BesselJ_0(\cdot)$ is Bessel function of the first order of index
0, and $\jr\approx3.8317\dots$ is the first positive root of its derivative,
$\BesselJ'_0(\jr)=0$. 
This problem has an exact solution:
\begin{equation}
  \u =\BesselJ_0\left(\jr \sqrt{(\x-\x_0)^2+(\y-\y_0)^2}\right)\,\e^{-\jr^2\t}.
                                        \eqlabel{HTestSolution}
\end{equation}
\Fig{discerror} illustrates the numerical convergence of the solution of
the problem~\eq{HTestPDE}--\eq{HTestBoundary} by \bbx\ using \code{diff} device to the exact solution
provided by~\eq{HTestSolution}, for $\T=0.2$. 
The timestepping is by using a
forward Euler scheme with a time step of $\dt=\dx^2/80$. The error, i.e. the difference
$\err(\r,\t)=\u\num(\r,\t)-\u(\r,\t)$ 
between the exact solution $\u(\r,\t)$ and its approximation obtained
numerically, $\u\num(\r,\t)$, is characterized by two
norms,
\begin{align}
  \Norm{\err}_{\Linf} = \max\limits_{\t\in[0,\T]}\max\limits_{\r\in \Domain}\Abs{\err(\r,\t)},
                                        \eqlabel{Linf}
\end{align}
\begin{align}
  \Norm{\err}_{\Ltwo} = \left(
  \frac{1}{\T\mes{\Domain}}
  \int\limits_0^\T\int\limits_{\Domain} \Abs{\err(\r,\t)}^2\,\d^2\r\,\d{\t}
  \right)^{1/2} .
                                        \eqlabel{Ltwo}
\end{align}
where $\mes{\Domain}$ stands for the area of $\Domain$, and
all the integrals are calculated by the trapezoidal rule. Each
$\dx$ is represented by four points on each graph, corresponding to
four simulations, with different position of the centre of the circle
$(\x_0,\y_0)$ with respect to the grid $\left(\r_\p\right)$,
namely, $\left((\x_0,\y_0)-\r_\p\right)/\dx=(0,0)$,
$(0.2,0.2)$, $(0.2,0.6)$ and $(0.6,0.6)$; this is to eliminate any
possible effects related to special arrangement of the problem with
respect to the grid.
We can
see that the convergence is worse than $\dx^2$, but better than
$\dx^1$. The $\Ltwo$ norm of the error converges faster than $\Linf$
norm, which is an indication that the main source of error is
localized --- this is, of course, to be expected, as the boundary
conditions, in a sense, approximate the curvilinear boundary
$\Boundary$ with pieces of straight lines
parallel to the $x$ and $y$ axes, thus typically making an error
$\O{\dx}$. We stress that in cases where the realistic tissue geometry
is available as a set of points with the same resolution as the
computational grid, the knowledge of any curvilinear boundary is in
any case unavailable, so any loss of accuracy associated with it, or,
equivalently, any notional gain of accuracy that would be associated
with using a curvilinear boundary instead, would be purely
theoretical.

\begin{figure}[htb]
\centering\includegraphics{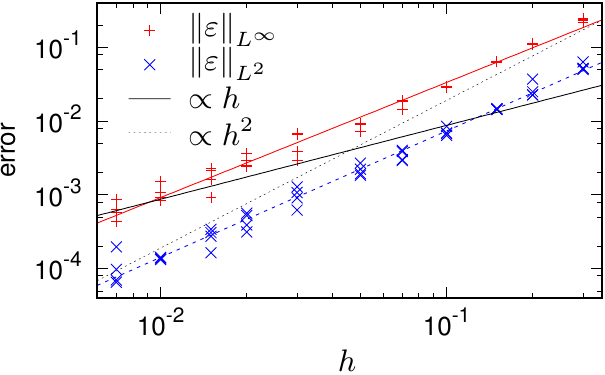}
\caption[]{%
{\bf
  Numerical convergence of the solution of the
  problem~\eq{HTestPDE}--\eq{HTestBoundary}.
}
  Slope lines are
  with slopes 1, 2 and best fits with slopes $1.564$ for
  $\Linf$, and $1.719$ for $\Ltwo$. %
}
\figlabel{discerror}
\end{figure}

From the MPI viewpoint, the work of the \code{diff} device and
  other similar devices requires special care, since computation of
  the Laplacian of a dynamical field at a point requires knowledge of
  the field in neighbouring points, and some of the neighbouring
  points may be allocated to different threads. Hence, some message
  passing (exchange of interfacial buffers) is required for its
  work. This issue is discussed in detail in~\secn{partitioning}.

\subsection*{Bidomain diffusion}
\seclabel{bidomain}

Computations using the bidomain tissue description~\eq{bidomain}
differ primarily by the presence of the equation:
\begin{align}
  \nabla\cdot(\sigi + \sige) \nabla\phii 
  = \nabla\cdot\sige \nabla\V - \Ie , \eqlabel{elliptic}
\end{align}
which is elliptic with respect to both $\phii$ and $\V$.
We have implemented a solver for elliptic equations in the 
\code{elliptic} device. This uses Full Multigrid iterations with
vertex-centered restriction/prolongation operators with bi/tri-linear
interpolation, and a (multicoloured) Gauss-Seidel or a Jacobi
smoother. 

The
linear system to which the solver applies naturally occurs through
discretization of the diffusion operator in the same way as described
in the previous subsection.
For solving the bidomain model~\eq{bidomain} using
operator splitting, the \code{elliptic} device can be used to solve the elliptic
equation with respect to $\phii$\cite{pinning}, leaving the parabolic diffusion
equation for timestepping $\V$ using the $\code{diff}$ device and
timestepping $\V$ and $\g$ according to the reaction kinetics via an
ODE solver, such as \code{euler} device.

The MPI aspect of the \code{elliptic} device is similar to that
  of \code{diff} device, in that the Gauss-Seidel iterations involve
  neighbouring grid points which may be allocated to different
  threads, and this is considered in detail
  in~\secn{partitioning}. The specifics of the \code{elliptic} device is that
  it implements an iterative algorithm, and buffer exchange
  is required at every iteration.  

We illustrate this computation scheme on an example of a bidomain
problem with an exact solution. We consider a
bidomain system~\eq{bidomain} with a one-component ``cell model'',
$\dim\g=0$, corresponding to the
Zeldovich-Frank-Kamenetsky~\cite{ZFK-1938} also known as Nagumo
equation~\cite{McKean-1970} and Schl\"ogl model~\cite{Schloegl-1972}:
\begin{align}
  & \df{\V}{\t} = \V(\V-\zfa)(1-\V) 
    + \left(\Dipar\pd^2_\x+ \Diort \pd^2_\y\right) \phii , 
                                        \eqlabel{ZFKbidomV}
  \\
  & \left[
    (\Depar+\Dipar)\pd^2_\x + (\Deort+\Diort)\pd^2_\y 
    \right]\phii
  = \left(\Depar\pd^2_\x +  \Deort\pd^2_\y\right)\V. 
                                        \eqlabel{ZFKbidomP}
\end{align}
If posed on the whole plane, $(\x,\y)\in\Real^2$, this system has a
family of exact solutions in the form of plane waves,
\begin{align}
  & \Va = \left\{1+\exp\left[\left(\x\cos\ang + \y\sin\ang - \s -
        \c\t\right)/\sqrt{2\Deff}\right]\right\}^{-1},
                                        \eqlabel{ZFKsolnV}
  \\
  & \phiia = \ptv \Va , 
                                        \eqlabel{ZFKsolnP}
\end{align}
where the angle of the wave propagation, $\ang$, and its initial
phase, $\s$, are arbitrary constants, and the other parameters of the
solution are defined by
$\c = \sqrt{2\Deff}\left(\frac12-\zfa\right)$, 
$\Deff = \Dieff\Deeff/(\Dieff+\Deeff)$, 
$\ptv = \Deeff/(\Dieff+\Deeff)$,
$\Dieff = \Dipar\cos^2\ang + \Diort\sin^2\ang$, 
$\Deeff = \Depar\cos^2\ang + \Deort\sin^2\ang$. 

For testing \bbx\ as a bidomain solver, we consider the problem for
the system~\eqtwo{ZFKbidomV,ZFKbidomP} in a square domain of size $\Len$,
$\Domain=[0,\Len]^2$, for a time interval $\t\in[0,\T]$, with the
initial and (non-homogeneous Dirichlet) boundary conditions set in
terms of~\eqtwo{ZFKsolnV,ZFKsolnP} as
\begin{align}
  & \V(\x,\y,0) = \Va,\quad \phii(\x,\y,0) = \phiia, \quad (\x,\y)\in\Domain; 
                                        \eqlabel{ZFKIC}
  \\
  & 
    \V(\x,\y,\t) = \Va, \quad 
    \phii(\x,\y,\t) = \phiia, \quad
    (\x,\y)\in\Boundary, \t\in(0,\T), 
                                        \eqlabel{ZFKBC}
  \\
  & \Boundary=\pd\Domain=\{0,\Len\}\times[0,\Len]\cup[0,\Len]\times\{0,\Len\}.
                                        \eqlabel{ZFKBoundary}
\end{align}

To implement solution of the
  problem~\eq{ZFKbidomV}--\eq{ZFKBoundary} in a \bbx\ script, we
split it into substeps:
\begin{align*}
\code{diff(1)}:  \quad & S_1 = \nabla\cdot \Deten \nabla \V, \\
\code{elliptic}: \quad & \nabla\cdot \left(\Deten+\Diten\right) \nabla\phii = S_1, \\
\code{diff(2)}:  \quad & S_2 = \nabla\cdot \Diten \nabla \phii,  \\
\code{euler}:    \quad & \V_\t = \V(\V-\zfa)(1-\V) + S_2. \\
\end{align*}
The resulting fragment of \bbx\ script looks like this: 
\begin{listing}
def str domain x0=xil x1=xir y0=yil y1=yir;
// source term in the elliptic equation
diff [domain] v0=[u] v1=[s]  
  Dpar=Dex Dtrans=Dey hx=hx;
// solving the elliptic equation
elliptic [domain] v0=[s] v1=[p] 
  Dpar=Dex+Dix Dtrans=Dey+Diy hx=hx  
  tolerance=tol delta=0.5 upper_level=3
  vcycles=20  preiter=1 postiter=2  maxiter=1e6;
// source term in the parabolic equation
diff [domain] v0=[p] v1=[s] 
  Dpar=Dix Dtrans=Diy hx=hx;
// timestep of the parabolic equation
euler [domain] v0=[u] v1=[u] 
  ode=zfk ht=ht par={alpha=alpha Iu=@[s]};
\end{listing}
In this fragment,
the first \code{diff} device computes the right-hand
side of the elliptic equation~\eq{ZFKbidomP}, $S_1=\nabla\cdot \Deten \nabla \V$,
and deposits the auxiliary variable  $S_1$ into the layer \code{[s]}; then
\code{elliptic} device solves the elliptic equation ~\eq{ZFKbidomP} for $\phii$, using the
provided fine-tuning algorithm parameters, and puts the result into the
layer \code{[p]}. The second \code{diff} device computes
$S_2=\nabla\cdot\Diten\nabla\phii$ and puts the result $S_2$ into the
layer \code{[s]} (which is therefore ``recycled''), and the
\code{euler} device does the time step of the cell model.
The interior of the domain $\Domain$ is mapped to the subgrid
\code{[domain]} with the 
grid $x$-coordinate from \code{xil} to \code{xir} and $y$-coordinate
from \code{yil} to \code{yir}.
The non-homogeneous, non-stationary Dirichlet boundary conditions
\eq{ZFKBC} were implemented by a \code{k\_func} device, computing the
boundary values for $\V$ and $\phii$ for the grid nodes surrounding
this subgrid \code{[domain]}, i.e. those with
grid coordinates \code{xil-1},\code{xir+1},\code{yil-1},\code{yir+1}
(this part of the script is not shown).

The accuracy of the computational scheme is illustrated
in~\fig{zfkerr}. We take $\Len=10$, $\T=40$, $\alpha=0.13$,
$\Dipar=2$, $\Diort=0.2$, $\Depar=8$, $\Deort=2$ and $\s=-5$.  The time step $\dt=3\dx^2/\left(16\Depar\right)$ is
varied together with the space step $\dx$
with
the coefficient $3/\left(16\Depar\right)$ chosen from the
  considerations of numerical
  stability~\cite{Clayton-Panfilov-2007-PBMB},
resulting in quadratic convergence
of the algorithm, as should be expected.
\begin{figure}[htb]
\centering\includegraphics{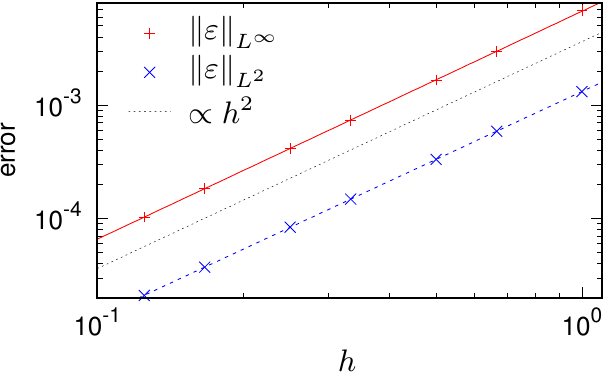}
\caption[]{%
  {\bf Numerical convergence of the solution of the
  problem~\eq{ZFKbidomV}--\eq{ZFKBoundary}. } Slope lines:
  slope 2 (black) and best fits with slopes $2.009$ for
  $\Linf$, and $1.9889$ for $\Ltwo$. %
}
\figlabel{zfkerr}
\end{figure}

\subsection*{Complex geometries and domain partitioning}
\seclabel{partitioning}

\subsubsection*{Format of \bbx\ geometry files}

Complex geometries for \bbx\ simulations are defined by files in \bbx's
``.bbg'' format. A \bbg\ file is an ASCII text file, each line in
which describes a point in a regular mesh. Each line contains
comma-separated values, in the following format: %
\[
 \code{ x , y , z , status , fibre\_x , fibre\_y , fibre\_z}
\]
Here \code{x, y, z} are integer Cartesian coordinates of the point,
\code{status} is a flag, a nonzero-value of which shows that this
point is in the tissue, whereas a zero \code{status} designates a
point in the void, and \code{fibre\_x, fibre\_y, fibre\_z} are $x$-,
$y$- and $z$-components of the fibre orientation vector at that
point. To reduce the size of \bbg\ files, only tissue points,
i.e. points with nonzero \code{status} need to be specified. \bbx\
will ignore the fibre orientation vectors of void points in any case.

\subsubsection*{Anatomically realistic geometries}

To use DT-MRI anatomical data in \bbx\ simulations, such data
should be converted into the \bbg\ format. The current version of
\bbx\ makes use only of the primary eigenvector of the diffusion
tensor, which is why only one direction vector is used in the format.
Once DT-MRI data on tissue points locations together with the
corresponding fiber orientations are compiled into a \bbg\ anatomy
file, it can be called from a \bbx\ simulation script, see {\it e.g.}
the statement 
\[
\code{state geometry=ffr.bbg anisotropy=1 vmax=4;}
\]
in \code{sample.bbs} script in \tab{sample} in the Appendix.

\subsubsection*{Domain partitioning}

Sharing work between processes in the MPI version of \bbx\ is done by
splitting the computational grid into \textbf{subdomains}, such that
computation in each subdomain is done by a process. The method of
domain decomposition is illustrated in~\fig{domdec}, for a 2D domain. Each of
the $\x$, $\y$ and $\z$ dimensions of the grid is separated by a certain
number of equal subintervals (approximately equal, when the grid size
is not divisible by the number of subintervals). The number of
subintervals in different dimensions do not have to be the same.
In the example shown in~\fig{domdec}, the $\x$ and $\y$
dimensions are split to have 3 subintervals each; the $\z$ dimension is
not split.  The grid nodes, in which computations are done, are
represented in~\fig{domdec} by solid circles (``bullets''). 

\begin{figure}[htb]
\centering\includegraphics{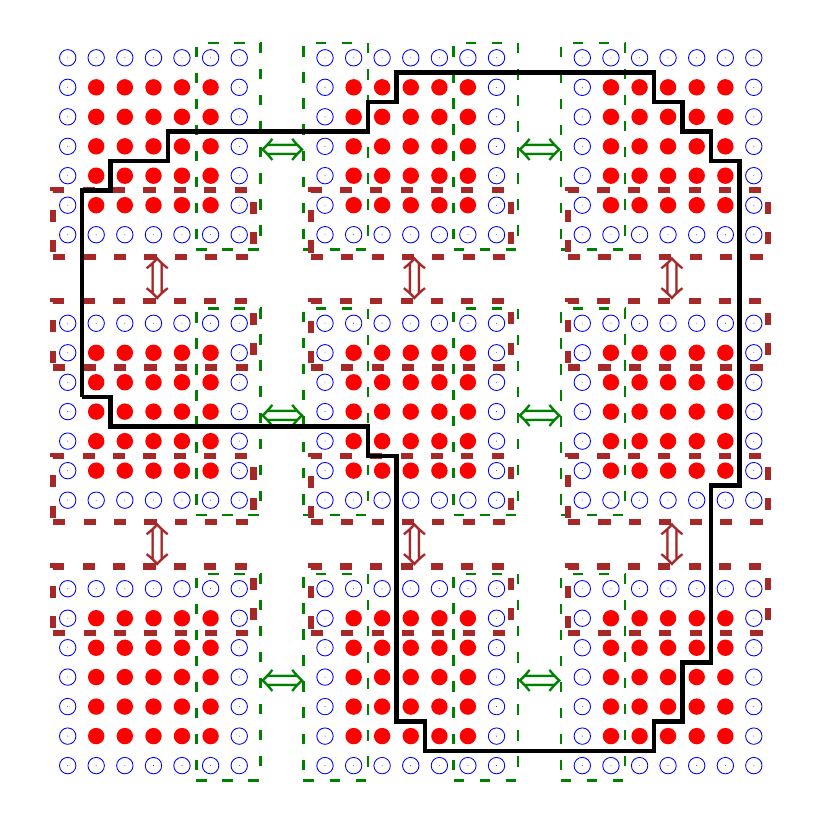}
\caption[]{%
 {\bf  Schematic of the domain partitioning in MPI implementation of 
  \bbx.
} Solid circles represent nodes on which actual computations
  are done, empty circles are the ``halo'' points, the rectangles
  denote the exchange buffers and the solid black line represents the
  boundary of an irregular computational domain (excitable tissue). %
}
\figlabel{domdec}
\end{figure}

The continuity of computations across subdomains necessary for 
devices involving the diffusion operator is achieved by using message
passing with \textbf{exchange buffers}. The depth of the exchange buffer
in each direction is one grid point. This imposes a limitation on the
stencils that can be used by diffusion-like devices, such as a 9-point stencil for
2D and up to a 27-point stencil for 3D. In~\fig{domdec}, the hollow circles
represent the fictitious grid nodes which are images of corresponding
nodes from neighbouring subdomains, and the dashed lines designate the
whole buffers, including the nodes to be sent and nodes to be
received. The buffer exchange should be effected twice (forwards and
backwards) for each dimension, i.e. four exchanges in 2D simulations
and six exchanges in 3D simulations. If the buffer exchanges are done
in the correct order, then this will ensure correct exchange of node
values in the diagonal directions as well (\textbf{magic corners}, see
~\cite{McFarlane-2010} for details).

In bidomain computations, the buffer exchanges can be done either
  before each iteration or more seldom; in the latter case the result
  is different from the sequential run, but inasmuch as both MPI and
  sequential results are close to the actual solution, the difference
  between the two should be negligible.  Similar consideration applies
  to the use of the Gauss-Seidel smoother, which of course will also
  give different results if applied in each subdomain separately.  

When working with complex (non-cuboidal) domains, the \bbx\ approach is
to inscribe the domain into the smallest cuboid, and then proceed as
before, with the difference that computations are only done at those
grid nodes that belong to that domain, and the one outside domain
remain idle. This is also illustrated schematically in~\fig{domdec},
where the boundary of the irregular domain is drawn by a closed bold solid
black line. This creates a challenge to the performance: with
high-degree parallelization and complicated geometry of the  irregular domain,
the load imbalance between processes can become significant; in
particular, a large number of partitions will contain no points of
the domain (in~\fig{domdec}, there is one such partition, in the
bottom left corner). This problem is well known and there
are efficient tools for solving them for structured as well as
unstructured grids, see e.g.~\cite{LaSalle-etal-2015,metis}. In the
current version of \bbx, however, only the crudest optimization method
is used: the partitions that are completely idle are not allocated to
processes, which considerably limits the expected slow-down because
of the uneven load (roughly speaking, at worst twice on average).

\subsection*{Input/output and visualization}
\seclabel{io}

Finally, we briefly mention some input/output options currently
available in \bbx. Each of these options is implemented in the
corresponding device. This includes:
\begin{itemize}
\item Full precision binary input and output of a specified subset
  grid, by the devices \code{load} and \code{dump}
  respectively. In the MPI mode, these devices
      operate in parallel, using \code{MPI\_File\_read\_all} and
      \code{MPI\_File\_write\_all} calls respectively. All
      threads get collective access to the file using
      \code{MPI\_File\_set\_view} calls at the start-up time, taking
      into account (computing the intersection of) both the 4D subset
      of the grid data allocated to the device, and the subdomain
      allocated for the particular thread. 
\item Discretized ``fixed-point'' (1 byte per value) output
of three selected layers of a subset of the grid, by the device
\code{ppmout}. MPI implementation of \code{ppmout}
   device is very similar to that of the \code{dump} device: use of 
      \code{MPI\_File\_write\_all} calls for writing after arranging
      collective access via 
      \code{MPI\_File\_set\_view} calls at the start-up time. The main
      difference is that \code{ppmout} device outputs only one byte
      per value instead of eight for \code{dump}. 
     Therefore \code{ppmout} precision is typically not sufficient
    for the \code{ppmout} output data
    to be used as control points or initial conditions for subsequent
    \bbx\ runs,
    though sufficient enough for most visualization purposes.
\item Plain text outputs of a defined subset of the grid
  by the device \code{record}, and a list of expressions involving
  global variables, by the device \code{k\_print}.
  The MPI aspects of the \code{record}  and \code{k\_print} devices are very
      different. The \code{record} device is similar to the
      \code{dump} and \code{ppmout} devices: it uses
      \code{MPI\_File\_set\_view} and \code{MPI\_File\_write\_all} for
      parallel output of a certain 4D subset of the grid 
 in a fixed-lengh ASCII format, so the position of each text
      record in the output file can be precisely calculated. The
      \code{k\_print} device outputs a specified list of values which
      may be expressions depending on global variables, which by
      definition are equally known to all threads. Hence the output is
      done only by one dedicated thread. 
All the MPI work needed here is selection of the communicating
      thread.  
\end{itemize}

Some computational devices also have i/o options. For instance, device
\code{k\_func}, which performs computations by formulas specified in
the \bbx\ script, can also read data from of a plain text file; such data
are interpreted as a tabulated univariate vector-function and is often
used to create initial conditions by the phase-distribution method~\cite{ft}.
This read in from a file is done via sequential calls, which
    in the MPI mode may create some delay, but since this is done only
    once at the start-up time, we do not consider this a significant
    issue. 

Another example is device \code{singz}, which finds phase
singularities in $z$-cross-sections of the grid. 
These are defined as intersections of isolines
  of two fields allocated to selected layers of the grid. Coordinates
  of the intersection points within grid cells are defined using
  linear interpolation of the pieces of the isolines.
In addition to
assigning the coordinates of the singularity points to global
variables, it can also output those to a plain text file and/or
visualize. %
The MPI implementation of the \code{singz} device is slightly more
    complicated than in other devices. The singularity points can be
    found in any threads, and their number is not known \textit{a
      priori}. Hence the coordinates of these points are passed, by
    \code{MPI\_Send} to one dedicated ``coordinating'' thread, which
    collates reports from all participating threads obtained via
    \code{MPI\_Recv}, and then assigns statistics of the found
    singularity points (their number, and means and standard
    deviations of their coordinates) and outputs, if requested, their
    coordinates to a file using sequential access. Naturally, the
    participating threads have to submit an empty report even if no
    singular points are found, as otherwise the coordinating thread
    has no means of knowing what messages to expect. This potentially
    creates an unnecessary delay compared to the sequential version;
    however in practice this is not noticeable since this device is
    usually called only during a small fraction of computation
    steps. %
Many devices have an optional \code{debug} parameter for printing
plain text messages about details of their work. %
The MPI versions of these options depend on
    whether the device operates with grid data or global variables,
    and is based on the same principles as described above.

Regarding run-time visualization, the sequential version of \bbx\ has
a number of devices for 0D, 1D, and 2D visualization 
via \code{X11} protocol if available (devices
\code{k\_draw}, \code{k\_plot}, \code{k\_paint}). 3D output typically
requires much more tuning in order to be effective. Theoretically, one
possibility is to do the tuning in the interactive mode while the
computations are stopped, as it is done e.g. in \code{EZSCROLL},
see~\cite{EZSCROLL}; this, however, would go against one of the
principles of \bbx, that all details of the run are specified in the
input script, so that any simulation is reproducible. Instead, currently
the 3D visualization is done by post-processing of the output data, most
often for the data produced by \code{ppmout}.

\section*{Results}

\subsection*{Parallel scaling performance}
\seclabel{scaling}

\Fig{scaling} illustrates the computation time taken by the MPI
version of \bbx\ on ARCHER (UK national supercomputing facility,
\url{http://www.archer.ac.uk/}) as a function of the number of
processes, for three series of test jobs, presenting different
challenges from the parallelization viewpoint.
In all series, the jobs simulated the monodomain model~\eq{monodomain}
with Courtemanche et al. 1998 model of human atrial
cells~\cite{Courtemanche-etal-1998}, with $\dim{\g}=23$, and
anisotropic diffusion (19-point stencil), but with different
geometries. %
\Fig{scaling}(a) is for a cubic grid of $300\times300\times300$
points. %
\Fig{scaling}(b) is for the rabbit ventricles geometry, described
in~\cite{Fenton-etal-RabbitHeart-2002}, which is inscribed in a cuboid
grid $119\times109\times131$, containing 470324 tissue points, which
is about 27.7\% of the total number of points in the grid. %
\Fig{scaling}(c) is for the human atrial
geometry~\cite{Seemann-etal-2006-PTRS}, 
inscribed into a
cuboid grid $237\times271\times300$, containing 1602787 tissue points,
about 8.3\% of the total.
The human atrium geometry tests were using the crude optimization of
partitioning, i.e. the subdomains that do not contain tissue points
are not allocated to processes; the rabbit ventricle jobs are done
without such optimization. 
In all job series, the simulation was with an initial condition of a
scroll wave with a filament along the $y$-axis, using the procedure
described e.g. in~\cite{Kharche-etal-2015-BMRI}. 
Simulation series for all three geometries were run with and without output
via \code{ppmout} device, which outputs up to three
selected layers of the grid, discretized to the 0..255 scale (3D
extension of the \code{ppm} format of Netpbm,~\cite{netpbm}),
using parallel output (see~\secn{io}
    for detail). The
corresponding curves in the graphs are marked as ``without ppm'' and
``with ppm'' respectively. Such outputs were done once in every 200
timesteps for the full box and human atrium geometries, and once in
every 1000 timesteps for the rabbit atrium geometry.
The full box and human atrium jobs also did plain text,
sequential output via \code{k\_print}
    device (see~\secn{io} for detail) of
the activity at a single point once in every 20 time steps.
To test the expenses associated with sophisticated control, the full
box and human atrium jobs implemented feedback-controlled stimulation,
similar to that implemented in the \code{sample.bbs} script described in the
Appendix.
To exclude the effects of the time taken by the start-up operations,
we computed the time per step by running jobs identical in all respect
except the number of time steps, and then considering the difference.
ARCHER has 24 cores per node, and the numbers of processes in the test
jobs are power-of-two multiples of 24.
The ``ideal'' lines are drawn based on the result of the
``without-ppm'' job on 24 processors. 

\begin{figure}[htb]
\centering\includegraphics{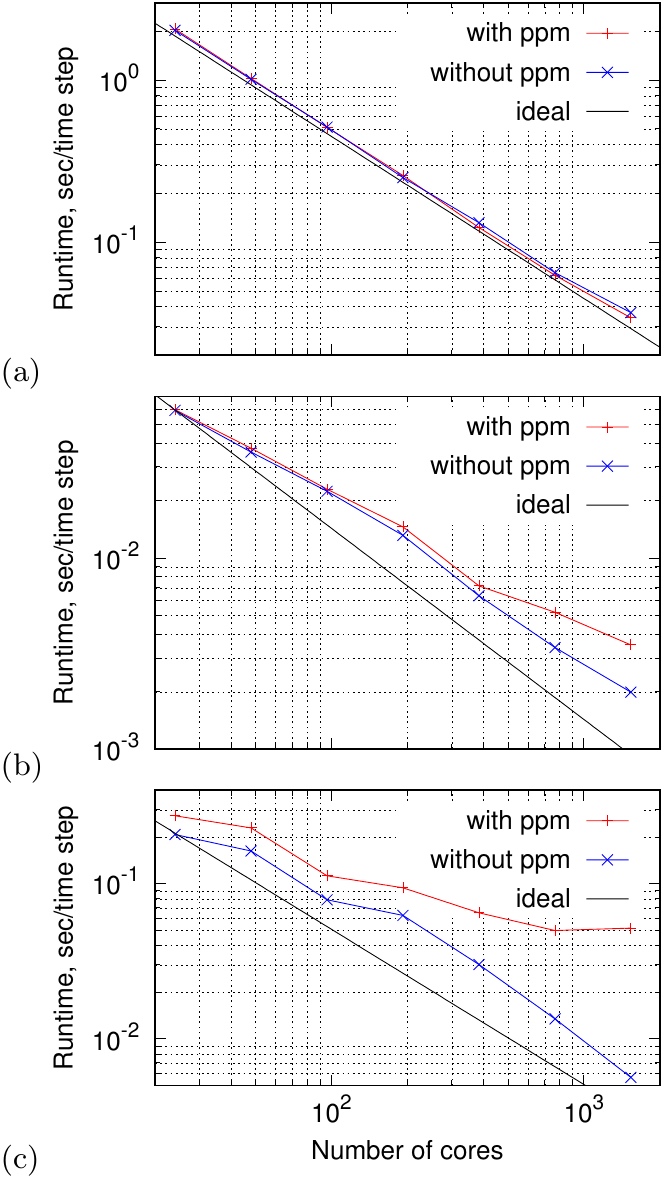}
\caption[]{%
  {\bf Log-log plots: the wall clock time per one time step in the
  simulation job, vs the number of
  cores. 
}%
  (a) Full box; %
  (b) Rabbit ventricle geometry, %
  (c) Human atrium geometry. %
  In all plots, ``with ppm'' stands
  for performance including file output via \code{ppmout} device,
  ``without ppm'' stands for pure computations, and ``ideal'' is
  the perfect-scaling extrapolation of the performance
  achieved on the smallest number of cores.
}
\figlabel{scaling}
\end{figure}

We observe that the effect of feedback control and plain text
outputs on the parallel performance is relatively small, and the main
slow-down at high parallelization happens due to uneven load of the
processes. The parallel scaling is, as expected, best for the full
box: without \code{ppmout} it is close to ideal for up to 1536
processes, while the curves for the complicated geometries deviate
from ideal noticeably. Human atrium geometry is proportionally ``much thinner'' than
the rabbit ventricle geometry, and the deviation from ideal is more
pronounced in~\fig{scaling}(c) than in \fig{scaling}(b). However, the
detrimental effect of the uneven load is limited: notice that the
``without ppm'' curve in \fig{scaling}(c) is almost parallel to the
ideal in the interval of 192--1536 cores, and the slow down is only
slightly more than by a factor of two. Another significant factor is
the bulk output. In the human atrium geometry, such outputs were 10
times more frequent, and their effects is more pronounced overall and
starts increasing at smaller parallelizations. Notice that the
relative effect of the bulk output is much less in the full box: an
obvious explanation is that the \code{ppmout} format always outputs
the full enclosing grid while computations are only done in the tissue
nodes, hence the output/computation ratio is about 12 times bigger for
human atrium than in the full box.
 
\Fig{scaling}(a) shows that on a standard test of full 3D
  box computation \bbx\  parallel performance is in accordance with
expectations and satisfactory. 
The maximal efficient parallelization is of
  course problem-dependent, as illustrated by \fig{scaling}(b) and \figref{scaling}(c).
  \Fig{scaling}(b) presents results of simulations of a small rabbit heart
  with less then $10^6$ grid tissue points, so at a larger number of
  cores, the inter-process communication expences take over the computation
  scaling. \Fig{scaling}(c) represents results of realistic simulations of a
  complex and ``thin'' human atrium, with less than $2\times10^6$ grid
  tissue points. 
  This illustrates the fact that parallel performance scaling depends
  on the ratio of inter-process communication costs to computations
  costs within one process, and for the cardiac modelling
  applications, this depends on the tissue geometry. We have verified
  that this ratio, and resulting perfromance, also
  depend on the excitation kinetic model: scaling is better if the
  kinetic model is computationally more complicated (these
  results are not shown). 
  Also, for the  simulations
with complex and ``thin''  geometries, a significant improvement may be
achieved by optimizing bulk outputs.

For the avoidance of doubt, we stress
    again that the performance results presented in~\fig{scaling} are
    per time step of simulation, and exclude the time spent on any
    start-up operations, such as parsing the \bbx\ script, reading the
    geometry file if used, domain partitioning if in MPI mode,
    tabulating the complicated functions if using ionic models,
    etc. This is done deliberately as these one-off operations
    typically take only few seconds at most altogether, and for
    realistic simulations are negligible.

\subsection*{Examples of use in recent and current research}
\seclabel{examples}

\bbx\ or its predecessors has been used to produce simulation results presented in dozens
of publications, e.g.~\cite{
      ft,%
      awt,%
      dew,%
      swd,%
      Biktasheva-etal-2015-PRL,%
      st,%
      Biktashev-etal-2011-PONE,%
      Kharche-etal-2015-BMRI,%
      Kharche-etal-2015-LNCS%
}.
In this section, we mention a handful of recent and
representative studies, illustrating the key features of this
software.

\Fig{ibz} illustrates a complicated
simulation set-up, which we believe would not be possible in
  other software currently available without re-coding at low-level
  requiring assistance from the developers.
This figure is an example
from~\cite{Biktashev-etal-2011-PONE} which modelled arrhythmogenic
mechanisms of the boundary layer between ischaemic and recovered cardiac tissue, moving due to
reperfusion. The model assumed that a certain ``excitability''
parameter 
(decrease in permeability of the inward potassium rectifier current $i_{K_1}$)
varied in space and time due to two factors: firstly,
space-only random distribution 
due to properties of individual cells;
secondly, deterministic smooth transition between low excitability of
the ischaemic tissue and high excitability of the recovered tissue,
changing in time due to reperfusion. On top of that, the
isotropic diffusion coefficient also varied along a similar
transition between low diffusivity of the ischaemic tissue and high
diffusivity of the recovered tissue, of a profile different from, but
moving synchronously with, the profile of the excitability parameter. 
\Fig{ibz} shows isosurfaces of the
transmembrane voltage field, $\V=-35\,\mV$, painted according to
the value of calcium current gating variable $f$, as shown by the
colourbar. In the snapshot shown, the upper half
of the box is a recovered, well connected, well excitable medium, which
supports a macroscopic scroll wave. The layer below it consists of
ischaemic boundary
cells that are less connected, so some of the cells (those where $i_{K_1}$
suppressed to a level below a certain threshold)  are in the
self-oscillatory regime
leading to a micro-scale turbulent
excitation pattern. The lowest layer consists of ischaemic cells with
suppressed excitability so they are not electrically active. As the
parametric profiles slowly move downwards, the solution represents
the process of recovery from ischaemia, which produces 
a reperfusion
arrhythmia as a result. All these spatio-temporal variations in
parameters have been set up not by writing special \code{C} code
for it, but at the \bbx\ script level using \code{k\_func}
device. Further detail can be found in
\cite{Biktashev-etal-2011-PONE}.

\begin{figure}[htb]
\centering\includegraphics{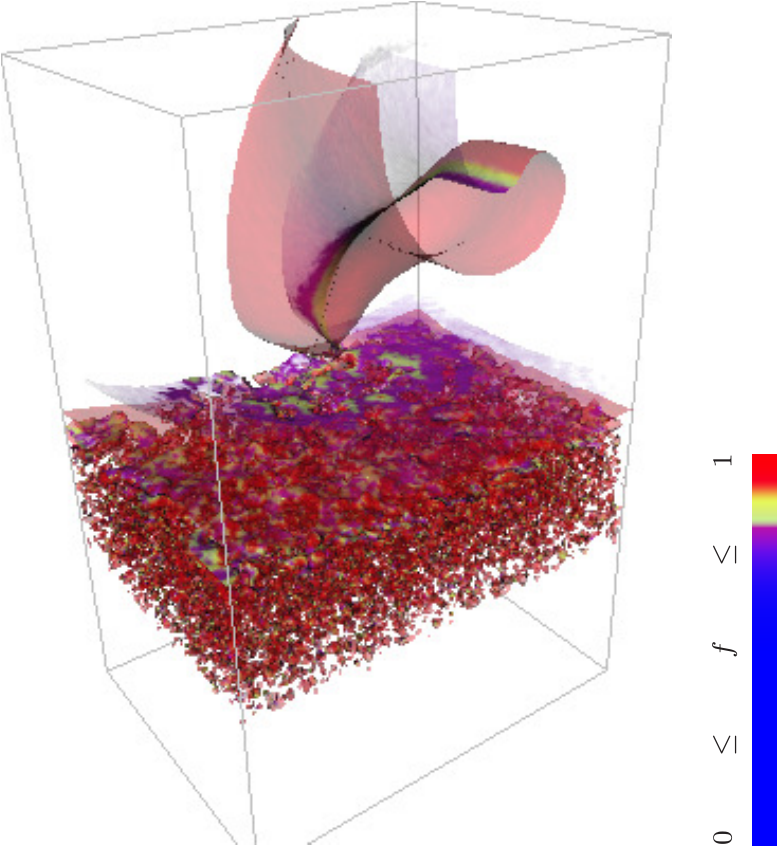}
\caption[]{%
{
\bf Scroll wave generation from ischaemic border zone.
}
  Generation of a scroll wave out of microscopic re-entries in
  excitable medium with random, space- and time-dependent distribution
  of parameters, modelling movement of ischaemic border zone during
  reperfusion~\cite{Biktashev-etal-2011-PONE}; Beeler-Reuter~\cite{Beeler-Reuter-1977} 
  kinetics.%
}
\figlabel{ibz}
\end{figure}
\Figs{dtf}, \figref{ha}, \figref{hv} illustrate simulations in
non-cuboid domains. \Fig{dtf} shows a surface view of a simulation in
an artificially defined domain, used to quantitatively test
predictions of an asymptotic theory about the drift of a scroll wave
in a thin layer due to sharp variations of thickness. This simulation
uses two-component FitzHugh-Nagumo kinetics.  Shown is the surface
view at a selected moment of time, colour coding represents states of
the activator variable (red colour component) and inhibitor variable
(green colour component) on dark-blue background, as
  shown by the ``colourbox'' to the right of the main picture; the
white line shows the trajectory of the tip of the spiral wave seen at
the upper surface of the domain for a period of time preceding 
the presented view. In this series
  of simulations, precise initial positioning of the scroll wave was
  required, which was achieved with the help of the ``phase
  distribution'' method implemented in \bbx's \code{k\_func} device.

\begin{figure}[htb]
\centering\includegraphics{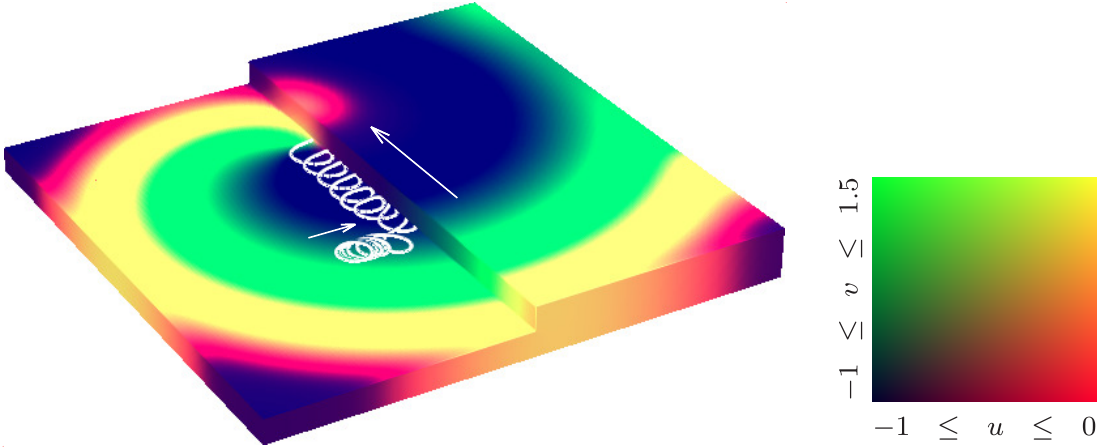}
\caption[]{
{\bf Drift  
  along a thickness step.
}
Drift of scroll wave 
  along a thickness step~\cite{Biktasheva-etal-2015-PRL},
  FitzHugh-Nagumo kinetics.}
\figlabel{dtf}
\end{figure}

\begin{figure}[htb]
\centering\includegraphics{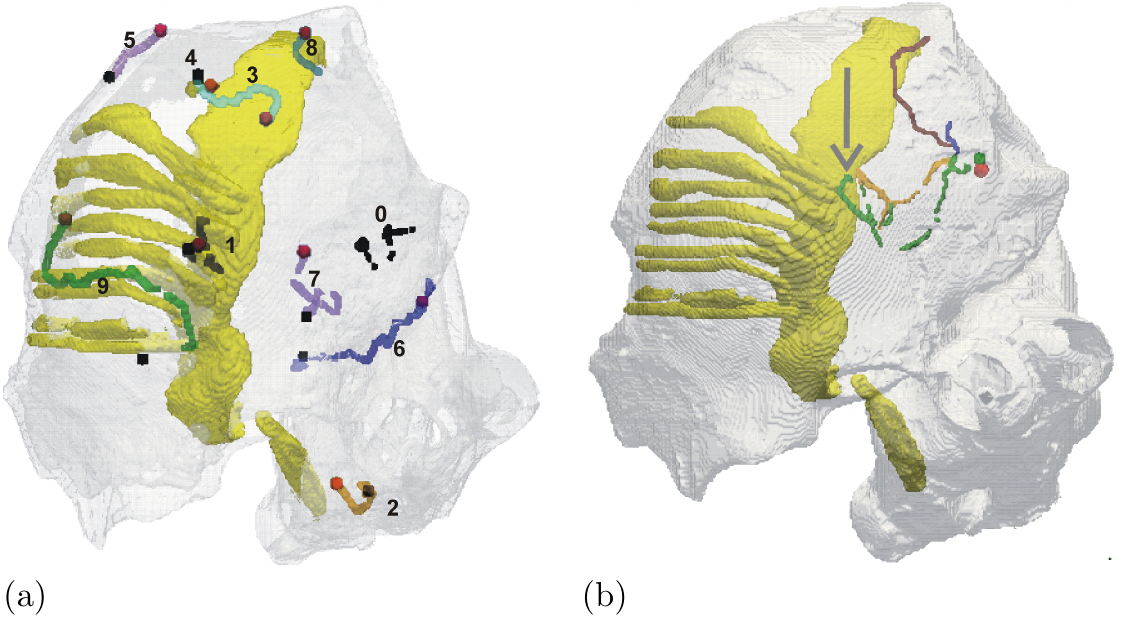}
\caption[]{%
{\bf Drift in a realistic human atrium geometry.
}
  Drift of scroll wave in a realistic human atrium
  geometry~\cite{Seemann-etal-2006-PTRS}, Courtemanche et al.~\cite{Courtemanche-etal-1998} kinetics.
  (a) Trajectories of spontaneous drift, caused purely by the anatomy features
  \cite{Kharche-etal-2015-BMRI}; %
  (b) Trajectories of resonant drift, caused by feedback-controlled electrical stimulation
  \cite{Kharche-etal-2015-LNCS}.%
}
\figlabel{ha}
\end{figure}

\begin{figure}[htb]
\centering\includegraphics{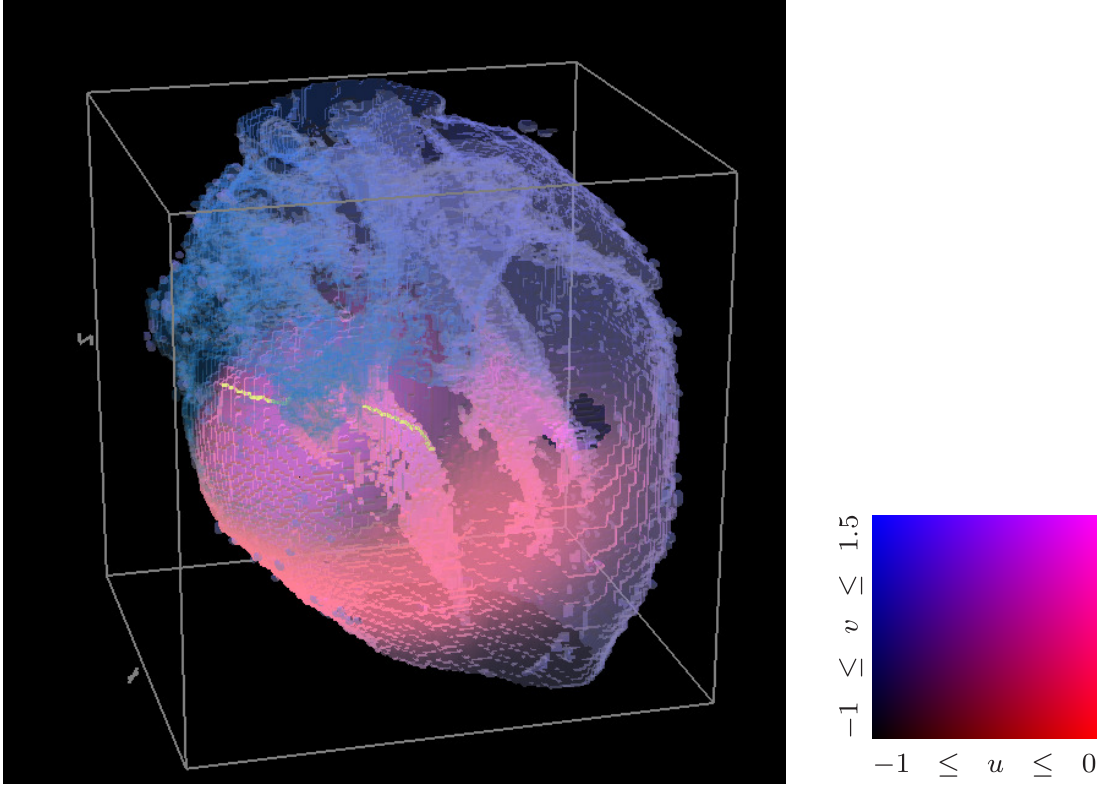}
\caption[]{%
{\bf Scroll waves of excitation in a DT-MRI based model of human foetal heart.
}
  A snapshot of excitation pattern with scroll wave
  filaments~\cite{Cai-etal-2016} in human foetal heart
  anatomy~\cite{Pervolaraki-etal-2013}, FitzHugh-Nagumo kinetics. 
  The surface of the heart is shown semitransparent,
  colour-coded depending on the values of the $\u$ and $\v$ variable
  as shown in the colourbox on the right. The yellow lines are the
  scroll filaments inside the heart.%
}
\figlabel{hv}
\end{figure}

\Fig{ha} illustrates simulations
    in an anatomically realistic model of human atrium, on a regular
    cuboidal MRI-type grid (although the actual grid
    origin was different, see \cite{Seemann-etal-2006-PTRS}).
\Fig{ha}(a) illustrates the anatomy-induced
drift~\cite{Kharche-etal-2015-BMRI}. 
Shown are a number of
trajectories of tips of spiral waves appearing on the surface of the
atrium nearest to the viewer; the yellow background indicates
prominent anatomical features (the pectinate muscles and the terminal
crest). To make the visualization clearer, the trajectories are
represented by lines connecting tip positions separated by exactly one
period of rotation (``stroboscopic view''); shown are several
trajectories starting at different initial positions and made within equal
  time intervals. Different trajectories are shown by
    different colours and enumerated. The beginning of each trajectory
  is shown by a red point, and the end of the trajectory is
    shown by a black
  point.

\Fig{ha}(b) illustrates further the \bbx's capability for
complicated simulation protocols. This panel displays results of
simulations in the same model as those shown in \fig{ha}(a), but now the
initial position for the scroll wave is chosen far from any sharp
features so that the anatomy-induced drift is not pronounced, and instead,
the scroll wave is subject to low-voltage pulses of external electric
field, $\Ieff=\Ieff(\t)$. The delivery of the pulses is controlled by
a feedback protocol similar to that illustrated by the sample
script~\tab{sample}, namely:
\begin{itemize}
\item There is a ``registration electrode'', at a point on the surface
  of the atrium that is the most distant from the viewer, position of
  which is indicated by the grey arrow. 
\item The signal from the registration electrode is monitored for the
  moment of arrival of an excitation wave, defined as the moment when
  the transmembrane voltage crosses a certain threshold value
  upwards. 
\item From the moment of the front arrival to the registration
  electrode, a certain waiting interval (delay) is observed. 
\item On expiration of the delay interval, a pulse of $\Ieff(\t)$ of a
  certain duration and certain amplitude is applied.
\end{itemize}
In \fig{ha}(b) we see three different trajectories starting at the
same point: they differ in the value of the delay interval between
registration of the front arrival and delivery of the electric
pulse. This stimulation protocol is aimed at achieving
 low-voltage defibrillation; the presented simulation
  illustrates the possibility of moving the location of a re-entrant
  arrhythmia by electrical stimuli of a magnitude much smaller than
  required for the classical single-shock defibrillation, in
    anatomically realistic setting, in different directions by adjusting
  the details of the feedback loop. This protocol is implemented
  fully at the \bbx\ script level, and we believe this would not be
  possible in any other software without re-coding at low level.

Both panels of \fig{ha} show tip trajectories
  starting at purposefully selected points; a specific challenge in
  this case was that for this particular study, it was important to
  have initial conditions that have only one phase singularity within
  the tissue, while the opposite one (necessary for topological
  reason) was about the big opening corresponding to the
  atrio-ventricular border. Again, the initial positioning of the
  scroll waves was done using the ``phase-distribution'' method,
  implemented with the help of the \code{k\_func} device at the \bbx\
  script level. Also, both series of simulations used ``stroboscopic''
  output, when the output data files were created in synchrony with a
  front of simulated excitation wave registered at a certain point of
  the medium; this also was implemented entirely at the \bbx\ script
  level.

\Fig{hv} from~\cite{Cai-etal-2016} shows a volume view of a scroll wave in a human foetal heart
geometry, 
obtained by DT-MRI~\cite{Pervolaraki-etal-2013}. 
Shown is the surface of the heart, semi-transparent
  and colour-coded depending on the values of the dynamic variables
  $\u$ and $\v$ of the FitzHugh-Nagumo kinetics \eq{fhn} chosen here for
  illustrative purposes.  The red component represents the $\u$ value,
  the blue component represents the $\v$ value, and the resulted
  colour-coding is summarised in the ``colourbox'' to the right of the
  main picture.  The yellow lines traversing the ventricular
  wall are the scroll filaments, defined as intersection of a
  $\u=\const$ surface with a $\v=\const$
  surface. 

The visualization in all cases was done by post-processing of the
simulation data. For \fig{ibz}, we used Iris
Explorer~\cite{IrisExplorer}. Both panels of \fig{ha} were generated
with ParaView~\cite{ParaView}. \Figs{dtf} and \figref{hv} were
produced by an in-house visualizer, based on the graphical part of
Barkley and Dowle's EZSCROLL~\cite{Dowle-etal-1997,EZSCROLL}, which is
in turn based on the Marching Cubes
algorithm~\cite{Lorensen-Cline-1987,Montani-etal-1994}.

\section*{Conclusion}
\seclabel{conclusions}

The leading idea underlying \bbx\ development is robustness,
portability, flexibility and user-friendliness in the first place,
connected with efficiency as an important but secondary
consideration. In the present form, \bbx\ can be exploited in
sequential and parallel (MPI) modes, with run-time and/or post-processing
visualization, on any unix-like platform from laptops to
supercomputing facilities. The modular structure of \bbx\ effectively
decouples the user interface, which at present is a scripting language
used to construct the ring of devices, from the implementation of the
computationally intensive stages in individual devices. The current
computational capabilities 
can be and will be further expanded in accordance
with the needs of wider usership without changing the backbone
ideology.

As far as MPI features are concerned, the straightforward approach to
parallelization via domain decomposition, yields acceptable results. 
As the maximal efficient parallelization is problem-dependent, small to medium scale anatomically
  realistic simulations become inefficient for number of threads
beyond about 1000. As higher-resolution DT-MRI anatomical
  data become available 
and/or more detailed kinetic
  models are used, the adequate parallelization should be expected to
  increase. However, \bbx\ already offers an important possibility of
  MPI utilisation of in-vivo MRI human heart anatomical data for real time
  simulations on multi-core desktop workstations for
  e.g. individualised ablation strategies, thus further broadening the
  MPI end users community. 

Apart from the size of the problem, another
important limiting factor is the uneven load of the parallel threads for ``thin'' complex
geometries of the computational domains, and output, which determines
possible direction of further development. The uneven load can be
addressed by a more careful fine-tuning of domain decomposition to
specifics of particular domain geometry, which to some extent may be
achieved without violating the main principles of the domain
decomposition, by allowing uneven partitioning along the coordinate
axes. 

The slow down in cases of extensive output is a problem which is
not specific for \bbx; however, some improvement in some cases may be
achieved by making any input-output operations exclusive to one or more
designated threads specializing on this and relieved from
computational load as such. 

As the current \bbx\ solvers use finite difference, regular
grid ideology, incorporation of DT-MRI regular cartesian grid anatomy
models into \bbx\ simulations is straightforward, as illustrated by
\figs{ha} and \figref{hv}, without a meshalizer
step required for finite element/finite volume solvers. However,
architecturally there is no
fundamental problems
in extending \bbx\ functionality to the finite element approximation as
long as regular mesh of finite elements is used that can be mapped to
a rectangular array. Extension to irregular meshes would require more
substantive changes, however the main idea of the ring of devices may
be useful there as well. The same concerns the
  ``discrete multidomain'' model of 
  in~\cite{Stinstra-etal-2006,Roberts-etal-2008}, which 
  describes cardiac tissue on the microscopic level. 
  Although one could attempt
  to embed this description into a regular grid, the most efficient
  implementation would require very different data structures.

Other relatively straightforward developments consistent with
  \bbx\ paradigm to be implemented in the foreseeable future,
  include: 
\begin{itemize}
\item Generalization of \code{diff} and \code{elliptic}
  devices for the orthotropic case.
\item Partitioning of the grid to domains described by
  different models. This can be used e.g. to model whole heart or
  its parts consisting of different tissues, surrounding bath or
  torso etc.
\item Fully automatic conversion of \code{CELLML} cellular
    models into the \code{rhs} format.
\item If and when the syntax of \code{CELLML} is enriched
    so as to explicitly identify gating and Markov-chain variables and
    their dynamics, fully automatic conversion of those into the
    \code{ionic} format.
\item Run-time 3D graphics (currently there is
    only 2D run-time graphics, and 3D visualization is done by
    post-processing).
\end{itemize}

\section*{Availability}

\bbx\ is free software available to download from the \bbx\ home page
\cite{beatbox-home}. The
source code is distributed under version 3 (or later) of the GNU
general public licence. %
The current version of \bbx\ software is also available in the S1 Code 
file.
\bbx\ is designed to be used in Unix-like operating
  systems, in non-interactive mode (started directly from the command line or by a
  shell script), with or without run-time graphics. The parallel version
  requires MPICH or OpenMPI, but the sequential version can be
  compiled and run without those. For the run-time graphics,
  \code{X11} is required, including its \code{GL} extension for some
  devices, but the computational part can be compiled and run without
  those.  Installation is done through the standard
  \code{configure}---\code{make}---\code{make install} command
  sequence, assuming that the environment includes \code{bash},
  \code{make} and a \code{C} compiler. Modifications to \bbx, such as
  adding new modules, would require the \code{autoconf} toolset.
  There are no other dependencies. Detailed installation instructions
  in \code{HTML} format are provided in the documentation supplied
  with the distribution~\cite{beatbox-home,beatbox-manual}, also in S1
  Code. The Matrix Rush-Larsen part of the
  \code{rushlarsen} device uses an eigenvalue solver from \code{GSL}
  library, but all relevant bits from \code{GSL} and its dependency
  \code{CBLAS} are included within the \bbx\ distribution, so the user
  need not worry about installing those separately, nor about the
  version compatibility.  

\section*{Supporting Information}

\paragraph*{S1 Code.}
\label{S1_Code}
{\bf \bbx\  software.} Zip-file containing
distribution of the \bbx, version beatbox-public-v1.7.982, including source code, configuration and makefiles, documentation, sample scripts etc.

\section*{Acknowledgments}
The human atrium DT-MRI data sets used in \BeatBox\ simulation presented in
\fig{ha} were provided by G.~Seemann et
al.~\cite{Seemann-etal-2006-PTRS}. The
human foetal heart DT-MRI data sets used in \BeatBox\ simulation presented in
\fig{hv} were provided by E.~Pervolaraki et
al.~\cite{Pervolaraki-etal-2013}. The
recent development of \BeatBox\ was supported by EPSRC grant
EP/I029664 (UK). VNB gratefully acknowledges the current
financial support of the EPSRC via grant EP/N014391/1 (UK).
IVB gratefully acknowledges the current financial support
of the EPSRC via grant  EP/P008690/1 (UK).
%
%
%

\appendix

\subsection*{Appendix: Examples of \bbx\  scripts}
\label{Appendix}
\subsubsection*{Script 1: \code{ez.bbs}}

\Tab{ezscript} provides a ``minimalist'' example of a
  \bbx\ script. It approximately emulates the functionality of
  Barkley's \code{EZSPIRAL}~\cite{ezspiral} (except tip finding and
  recording, saving the final state, and starting from a previously
  saved state).  Namely, it performs a simulation of the Barkley
  model~\cite{Barkley-1991}
  on a 2D grid consisting of $100\times100$ internal points;
  one extra row of points in each direction is added to implement the
  boundary conditions. The initial conditions are specified using
  ``instant cross-field'' protocol:
\begin{align*}
  \u=
  \begin{cases}
    1, & \y>\yc, \\
    0, & \textrm{otherwise,}
  \end{cases} 
  \qquad
  \v=
  \begin{cases}
    0.4, & \x>\xc, \\
    0, & \textrm{otherwise,}
  \end{cases}
\end{align*}
where $(\xc,\yc)$ is the centre of the box.  Every 10 time steps, it
plots the solution in an OpenGL window (using the colour-coding
similar to that of~\fig{dtf}), and outputs the dynamic variables into
a text file.

\begin{table}[htbp]
  \begin{listing}
/* Box of 100x100 internal points, 3 layers */
state xmax=102 ymax=102 vmax=3;
/* Schedule control flags */
def real begin;		// true only at the beginning
def real out;		// true when graphic and text outputs are due
def real end;		// true when all done
/* The schedule: this k_func computes only global variables, at each t */
k_func nowhere=1 pgm={begin=eq(t,0);out=eq(mod(t,10),0);end=ge(t,1000)};
/* Init. cond.: this k_func computes only local field values, at t=0 only */
k_func when=begin pgm={u0=gt(y,50); u1=0.4*lt(x,50)};
/* Graphic output of u and v fields distribution */
k_paintgl when=out width=300 height=300 nabs=100 nord=100
  pgm={red=u(abs,ord,0,0); grn=u(abs,ord,0,1)/0.8; blu=0};
/* Text output of a point record */
record when=out x0=10 x1=10 y0=20 y1=20 file="history.dat";
/* Terminate when all work done */
stop when=end;
/* Diffusion substep for layer 0, layer 2 reserved for Laplacian */
diffstep v0=0 v1=2 ht=0.02 hx=0.4 D=1;
/* Reaction substep for layers 0:1; Barkley's variation of FitzHugh-Nagumo kinetics */
euler v0=0 v1=1 ht=0.02 ode=fhnbkl par={a=0.8 b=0.01 eps=0.02};
end;
  \end{listing}
\caption[]{{\bf {\small \bbx\  script} \code{ez.bbs}.} A simple \bbx\ script}
\tablabel{ezscript}
\end{table}

The main features of the syntax may be seen from the
script itself which is intended to be self-explanatory, but
nevertheless:
\begin{itemize}
\item Comments in the script can be in C style, within \code{/*...*/} or in
  C++ style, between \code{//} and the end of line. 
\item The script is a sequence of sentences, each concluding with a
  semicolon, `\code{;}'.
\item Sentences starting with the keyword \code{def} declare global
  variables. 
\item The sentence starting with the keyword \code{state} allocates the
  computational grid. 
\item The script finishes with a sentence ``\code{end;}''.
\item Other sentences describe instances of devices comprising the
  ring. The first keyword in each sentence is the device type; other
  words describe the parameters determining the specifics of the work
  of this particular instance of the device.
\end{itemize}
The particular sentences in the script have the following functions: 
\begin{itemize}
\item \code{state} sentence, preceding any devices, defines and
    allocates the
  computational grid. In this case the space domain is a 2D box: the $z$-dimension
  is not specified so defaults to \code{zmax=1}. The parameter
  \code{vmax=3} means there will be three layers in the grid, 
  numbered 0,1 and 2. As we shall see, layer 0 is reserved for the $\u$
  field, layer 1 for the $\v$ field, and layer 2 is used for computing
  and storing the diffusion term. 
\item \code{k\_func}, the first device in the script, computes, depending on the
  current value of the loop counter \code{t}, the ``flag'' global
  variables that control which of the other devices will or will not
  work at the current time step iteration. As this device changes values
  of global variables, it is not allowed to change local field values,
  hence \code{nowhere=1} parameter. This instance of \code{k\_func}
  works at the beginning of every time iteration, and as a result,
  variable \code{begin} will take the value of 1 at the very first
  iteration and 0 otherwise; variable \code{out} will take value 1
  only when the loop counter \code{t} is divisible by 10, i.e. at
  every 10-th iteration, and variable \code{end} will become one as
  soon as the counter \code{t} exceeds 1000.
\item The second device in the script is another instance of
    \code{k\_func} device. Now
  it computes not the global variables, but the values of the
  field variables at every point of the space grid, according to the given
  formula. According to the \code{when=begin} parameter, this device
  works only once, at the very first time step, and its function is to
  produce initial conditions for the simulation.
\item \code{k\_paintgl} is a graphic output device.  It creates an X11
  window of $300\times300$ pixels,  and at every tenth timestep
  (according to the parameter \code{when=out}), paints using OpenGL
  a $100\times100$ raster, each element of which will be
  coloured according to the given formulas: the relative luminosity of
  the red component is equal to the value in layer 0 (corresponding to
  the $\u$ field), for the green component it is equal to the value in
  layer 1 (corresonding to the $\v$ field) divided by 0.8, and the
  blue component always is zero. Note that this colour-coding is similar to the
  the colour-coding used in~\fig{dtf}. 
\item \code{record} device opens for writing a text file
  \code{history.dat}, and at every tenths timestep (according to
  \code{when=out}), will print into the file the values of the grid nodes
  within the cuboid subdomain defined by the parameters \code{x0}
  \dots \code{v1}, which makes exactly two values: layer 0
  ($u$-field) and layer 1 ($v$-field)
  values at the point of the grid with integer coordinates $(10,20)$.
\item \code{stop} is the device whose function is to interrupt the
  computations and terminate the program. Naturally this device must
  be present in the ring unless it is intended that the program run is
  to be interrupted by the operator. In the presented example, the
  device works simply when the global variable \code{end} takes a nonzero
  value, which happens after 1000 time steps. 
\item \code{diffstep} is the first of the devices which does ``the
  actual computations'' in the sense that it changes the the  field
  variables in the layers of
  the computational grid
  according to the differential equations.
  As could be guessed from its
  name, it computes the sub-step due to the diffusion
  term. Specifically, it computes a value of the diffusion term, for the
  $\u$-field stored in layer 0 of the computational grid, using
   the given values of the diffusion coefficient \code{D} and space
   discretization step
  \code{hx},   places the computed Laplacian into layer 2 reserved for this purpose, and then performs a
  forward Euler step for the $\u$-field for the given value of
  the time step \code{ht}. 
\item \code{euler} is a computational device which performs the
  forward Euler step for the dynamic fields stored in layers 0
  and 1 of the computational grid, with account of the given kinetic
  model.
\end{itemize}

\subsubsection*{Script 2: \code{sample.bbs}}

\Tab{sample} presents the
complete listing of a more non-trivial example of a \bbx\
script, \code{sample.bbs}. This is the example represented by the ``device ring'' 
in~\fig{ring}.
Some new syntax features observed in the script include:
\begin{itemize}
\item Expression \code{<fhn.par>} means inclusion of an ASCII text file
  with name \code{fhn.par}, as part of the script, similar to
  \code{\#include <fhn.par>} in C.  On this occasion, the file \code{fhn.par}
  contains definitions of the global variables, which are intended to be
  model parameters shared between many related scripts.
\item The declarations of the global variables in the \code{def}
  sentences may specify optional initial values, which are
    allowed to be defined
  by arithmetic expressions with previously defined or pre-defined
  variables.
\item Declarations of global variables may appear not only in the very
  beginning, but throghout the script. The only restriction is that a
  variable has to be declared before it is used.
\item Global variables of type \code{str} are string macros.
  Expansion of a string macro declared as ``\code{def str foo bar;}'' is
  done using syntax \code{[foo]} which will produce \code{bar} in
  place of expansion.
\item Overall, the values of the model/simulation parameters are often specified by
  arithmetic or string expressions rather than literal values;
  moreover, string macro substitutions are used in the body of a
  device definition. For instance, since the string macro \code{u} is
  defined as \code{0}, expression \code{u[u]} expands to \code{u0},
  and since string macro \code{0} is predefined to the sciript name,
  \code{sample}, the expression \code{file=[0].rec} expands to
  \code{file=sample.rec}.
\item Some of the devices in the script have parameter \code{name}.
  This allows to distinguish between different instances
  of the same device in the diagnostic messages in the
  simulation's standard output
  and the log file.
\end{itemize}

\begin{table}[htbp]
\begin{adjustwidth}{-1.25in}{0in}
\begin{listing}
<fhn.par>                                         // model pars are read in from file fhn.par
def str u 0;					  // u field in 1st layer
def str v 1;					  // v field in 2nd layer
def str i 2;					  // diffusion term in 3rd layer
def str b 3;                                      // spatially dependent parameter in 4th layer
def real grad [1];                                // its gradient is 1st command-line parameter
// Integer and real stimulation parameters
def int xr 100; def int yr 100; def int zr 100;   // reg electrode position, in space steps
def int dr 5;                                     // reg elecrode size, in space steps
def real Amp 3.0;                                 // pulse amplitude
def real Dur 0.1;                                 // pulse duration
def real Del 6.0;                                 // pulse delay
def real Tstart 100.0;                            // when to switch on the feedback 
state geometry=ffr.bbg anisotropy=1               // the file contains tissue geometry and fibres
  vmax=4;                                         // 2 dyn vars + diffusive current + parameter
def real T;def real begin;def real out;def real end; // real vars control work of some devices
k_func name=timing nowhere=1 pgm={                // this function operates only global variables
  T     = t*ht;                                   // t is integer time counter; T is real time
  begin = eq(t,0);                                // 1.0 at the very beginning, otherwise 0.0
  out   = eq(mod(t,100),0);                       // 1.0 every 100 timesteps, otherwise 0.0
  end   = ge(T,100.0)};                           // 1.0 after 100 ms, otherwise 0.0
// This function operates at every space point but only at the first time step
k_func name=IC when=begin pgm={ 
  u[u]=ifle0(x-25,1.7,-1.7); u[v]=ifle0(y-25,0.7,-0.7) // Cross-field initial conditions
  u[b]=bet+grad*(z-0.5*zmax)};                      // vertical gradient of parameter
// The feedback 
def real signal;def real front; def real Tfront;
reduce  operation=max result=signal v0=[u] v1=[u] // signal=max of voltage field within given volume
  x0=xr xr1=xr+dr-1 y0=yr yr1=yr+dr-1 z0=zr zr1=zr+dr-1; // the values are arithmeitc expressions
k_poincare nowhere=1 sign=1                       // remember T when signal crossed value umid upward
  pgm={ front=signal-umid; Tfront=T };            
k_func name=feedback nowhere=1                                  // force lasts Dur ms starting Del ms after crossing
  pgm={ force=ht*Amp*ge(T,Tstart)*ge(T,Tfront+Del)*le(T,Tfront+Del+Dur) };
// The computation
diff v0=[u] v1=[i] Dpar=D Dtrans=D/4 hx=hx;       // anisotropic diffusion
k_func name=stim when=force pgm={u[i]=u[i]+force};          // this applies everywhere, only when force is nonzero
euler v0=[u] v1=[v] ht=ht ode=fhncub              // cubic FitzHugh-Nagumo kinetics
   par={eps=eps bet=@[b] gam=gam Iu=@[i]};        // varied beta and current as calculated before
// Output
ppmout  when=out file="[0]/%04d.ppm" mode="w"     // every 100 timesteps: 
  r=[u] r0=umin r1=umax                           //    value-discretized
  g=[v] g0=vmin g1=vmax                           //    output for subsequent
  b=[i] b0=0 b1=255;                              //    visualization
k_print when=always file=stdout list={T; force; signal}; // to monitor work of the feedback 
record  when=end file=[0].rec when=end v0=0 v1=1; // ascii dump of all field values in the end of run
stop when=end;
end;
\end{listing}
\end{adjustwidth}
\caption[]{{\bf {\small \bbx\ script} \code{sample.bbs}.} A more complicated \bbx\ script.}
\tablabel{sample}
\end{table}

The particular devices used in the script, in order of
  occurrence, have the following functions:
\begin{itemize}
\item \code{state} sentence defines a complex
  geometry, read from the file \code{ffr.bbg}. Further, the diffusion
  will be anisotropic (\code{anisotropy=1}),
    with the fiber directions read from the same
  file, \code{ffr.bbg}. 
\item The first instance of \code{k\_func},  with the name \code{timing}
  computes the ``flag'' global variables that control which of the
  other devices will or will not work at the current time iteration.
  Besides, it also computes the global variable \code{T}, which
  is to contain the model time $t$, as opposed to integer \code{t}
  which is the loop counter.
\item The second instance of \code{k\_func}, with the name \code{IC}
  computes the initial
  conditions. This time it computes not only $u$ and $v$ field
  allocated in layers \code{[u]} and \code{[v]}, but also the values
  of layer \code{[b]}, i.e. layer 3. The latter will contain not a
  dynamic variable, but values of the model kinetics parameter
  $b$,  which 
  in this simulation
  has a spatial gradient in the $z$ direction.
\item \code{reduce} is a device that computes the value of the global
  variable \code{signal} based on the current state of one or more of the
  fields represented in the layers of the computational grid; in this case it 
  uses just the
  layer \code{[u]}. 
  Here the \code{reduce} device emulates
  the work of a registration electrode, which measures the maximal
  value (parameter \code{operation=max}) of the ``transmembrane
  voltage'' in a particular small volume in the space grid, of the size
  \code{dr}$\times$\code{dr}$\times$\code{dr}, cornered at
  (\code{xr},\code{yr},\code{zr}). This measurement will be used as a
  feedback signal to control the electrical
  excitation in a putative low-voltage defibrillation protocol.
\item \code{k\_poincare} is a device that implements the idea of a
  Poincare cross-section from the dynamical systems theory.  It
  operates only global variables, hence does not have any domain
  associated with it, thus \code{nowhere=1}.  Here the \code{k\_poincare}
  device checks whether at the current iteration the signal,
  represented by variable \code{signal} measured by the previous
  \code{reduce} device, has crossed a given threshold value
  \code{umid} in the required direction, defined by \code{sign=1},
  which means upwards. If that has happened, then a certain flag (the
  global variable \code{front}) is ``raised'' (gets the values of 1),
  and the time, represented by \code{T}, when this happened is
  remembered in another global variable, \code{Tfront}.
\item The next instance of device \code{k\_func}, with the name \code{feedback},
  works with global variables: it
  computes, using the given formula, the value of the variable
  \code{force} that
  defines the defibrillating electric field, depending
  on the time that has passed since the event registered by the
  \code{k\_poincare} device at time \code{Tfront}, so that \code{T} is
  between \code{Tfront+Del} and \code{Tfront+Del+Dur}, where the variable
  \code{Del} is the delay of the stimulus compared to the front
  registration moment, and \code{Dur} is its duration.
\item \code{diff} is a computational device, which computes the
  diffusion term, i.e. the value of the Laplacian of the field
  represented by layer \code{[u]} of the computational grid, and
  records the result into layer \code{[i]} of the grid. Since the
  geometry defined by the \code{state} sentence is anisotropic, this
  \code{diff} device requires two diffusion coefficients, \code{Dtrans} and
  \code{Dpar} for conductivity across and along the fibers
  respectively. 
\item The next instance of \code{k\_func} device with the name \code{stim} is
  ``local'', i.e. it works on the computational grid: computes the
  action of the defibrillating electrical field (computed by the
  previous ``\code{feedback}'' instance of \code{k\_func}  device) onto the excitable cells. The
  action is simply adding the previously computed \code{force} to the
  layer \code{[i]}, which already contains the value of the diffusion
  term. 
\item \code{euler} here performs the time step
  for the dynamic fields in layers \code{[u]}..\code{[v]} represented in the
  computational grid, with the account of the given kinetic model
  \code{fhncub}, which is classical FitzHugh-Nagumo with cubic
  nonlinearity~\eq{FHN}. The new features here are the definitions
  of the ``extra current parameter'' \code{Iu} and of the parameter
  \code{bet} using symbol \code{@}. The meaning of this symbol is that
  the values of the parameter \code{bet} are taken from the layer
  \code{[b]}, that is layer 3, and the values of the parameter
  \code{Iu} are taken from the layer \code{[i]} defined as 2.
   This is a typical simplified mono-domain
  description of the action of the external electrical current, which
  in this simulation is assumed to be purely time-dependent,
  i.e. applied uniformly throughout the tissue.
\item \code{ppmout} output device works once
  in 100 steps (according to the computation of the \code{out}
  variable by the \code{timing} instance of \code{k\_func}) and
  produces an output to a file in \code{ppm} format, where each
  byte represents a value of one element of the grid, from up to three
  selected layers of the grid, discretized to the 0..255 scale. This
  \code{ppm} image format could be converted to other
  popular and less space-consuming formats either by postprocessing or
  on-the-fly (not done in the current \code{sample.bbs} script). The name of the
  \code{ppm} output file contains the \code{\%} symbol, the effect of
  which is that it is the format of a \code{C} \code{sprintf} call,
  the first field argument of which is the ordinal number of the
  device's instant call. That is, this device will produce files with
  names \code{sample/0000.ppm}, \code{sample/0001.ppm},
  \code{sample/0002.ppm} etc.
\item \code{k\_print} is a more straightforward output device: each
  time it is called (here, at the every time step), it
  adds to the output file a plain-text record of the values of the global
  variables involved in the feedback control of the
  defibrillating stimuli. It is similar to the \code{record} device
  in the \code{ez.bbs} above, except it prints global variables
  rather than grid node values.
\item \code{record} is the last output device in this script. Its use
  in this script is different from that in \code{ez.bbs}, in that it
  prints the values of the field layers 0 and 1 in all internal points
  of the grid. This device works only at the last time step of the
  simulation, so that the output file can be used as an initial condition
  if continuation of the present simulation is required.
\item \code{stop} is the last device in this script and its syntax
  and semantics is the same as in \code{ez.bbs}. 
\end{itemize}


\begin{thebibliography}{10}
\bibitem{bbx-early}
The early stages of development of \BeatBox\ and its predecessor QUI benefited
  from contributions by A.V.~Karpov and R.~McFarlane. Karpov provided the
  portable compiler of arithmetic expressions for QUI. McFarlane contributed to
  MPI parallelization of QUI, and treatment of complex geometries, and is the
  author of its new name, \bbx.
\end{thebibliography}
\end{document}